\documentclass[12pt]{article}
\usepackage{epsfig,amsfonts,amssymb}
\usepackage{cite}

\def\ZZZ{{\hbox{ Z\kern-1.6mm Z}}}
\def\zzz{{\hbox{ z\kern-1mm z}}}

\newcommand{\BB}{{\cal B}}

\newcommand{\HH}{{\cal H}}

\newcommand{\OO}{{\cal O}}

\newcommand{\EE}{{\cal E}}
\newcommand{\LL}{{\cal L}}

\newcommand{\wt}{\widetilde}
\newcommand{\wh}{\widehat}

\newcommand{\SSS}{{\cal S}}

\newcommand{\be}{\begin{equation}}
\newcommand{\ee}{\end{equation}}
\newcommand{\ben}{\begin{eqnarray}\displaystyle}
\newcommand{\een}{\end{eqnarray}}
\newcommand{\refb}[1]{(\ref{#1})}
\newcommand{\p}{\partial}

\renewcommand{\theequation}{\thesection.\arabic{equation}}
\def\one{{\hbox{ 1\kern-.8mm l}}}
\def\zero{{\hbox{ 0\kern-1.5mm 0}}}

\newcommand{\bea}{\begin{eqnarray}}
\newcommand{\eea}{\end{eqnarray}}
\newcommand{\beas}{\begin{eqnarray*}}
  \newcommand{\eeas}{\end{eqnarray*}}

\def\scriptlap{{\kern1pt\vbox{\hrule height 0.8pt\hbox{\vrule width
        0.8pt \hskip2pt\vbox{\vskip 4pt}\hskip 2pt\vrule width
        0.4pt}\hrule height 0.4pt} \kern1pt}}

 \def\Biggg#1{{\hbox{$\left#1\vbox to
        25pt{}\right.\n@space$}}} \def\n@space{\nulldelimiterspace=0pt
  \m@th} \def\m@th{\mathsurround = 0pt}

\def\xv{{\bf x}}
\def\Fv{{\bf F}}

\def\bbR{{\bf R}}
\def\eqref#1{(\ref{#1})}
\def\tfrac#1#2{{\textstyle\frac{#1}{#2}}}

\def\ap{\alpha'}

\makeatletter
\def\eqnarray{%
  \stepcounter{equation}\def\@currentlabel{\p@equation\theequation}%
  \global\@eqnswtrue \m@th \global\@eqcnt\z@ \tabskip\@centering
  \let\\\@eqncr
  $$\everycr{}\halign to\displaywidth\bgroup
    \hskip\@centering$\displaystyle\tabskip\z@skip{##}$\@eqnsel
   &\global\@eqcnt\@ne \hfil$\;{##}\;$\hfil
   &\global\@eqcnt\tw@ $\displaystyle{##}$\hfil\tabskip\@centering
   &\global\@eqcnt\thr@@ \hb@xt@\z@\bgroup\hss##\egroup
    \tabskip\z@skip
    \cr}
\makeatother

\makeatletter
\@addtoreset{equation}{section}
\makeatother

\begin{document}

%

\baselineskip=18pt

\begin{titlepage}
  \begin{flushright}
   {\small TIFR/TH/06-36},
   {\small NSF-KITP-06-104} \\
   {\small CALT-68-2617},
   {\small HRI-P-06-11-001}\\
   {\small hep-th/0611166}
  \end{flushright}

  \begin{center}

    \vspace{5mm}
  {\Large \bf Spinning Strings as Small Black Rings} \\

 \vspace{6mm}
    $^{1}$Atish Dabholkar, $^{2}$Norihiro Iizuka, $^{1}$Ashik Iqubal, \\
    $^3$Ashoke Sen
    and $^4$Masaki Shigemori
    \vspace{3mm}

    {\small \sl $^1$Tata Institute of Fundamental Research\\ Homi Bhabha
    Road, Mumbai, 400 005, INDIA} \\
   \vspace{3mm}
    {\small \sl $^2$Kavli Institute for Theoretical Physics, \\
    University of California, Santa Barbara, CA 93106-4030, USA }\\
   \vspace{3mm}
    {\small \sl $^3$Harish-Chandra Research Institute\\ Chhatnag Road,
    Jhusi, Allahabad 211019, INDIA}\\
   \vspace{3mm}
    {\small \sl $^4$California Institute of Technology 452-48,
    Pasadena, CA 91125, USA}\\
   \vspace{3mm}
    {\small \tt atish@theory.tifr.res.in, iizuka@kitp.ucsb.edu, 
iqubal@tifr.res.in, \\
    sen@mri.ernet.in, shige@theory.caltech.edu}
   \vspace{3mm}

  \end{center}
%
  \noindent

Certain  supersymmetric elementary string states with spin can be
viewed as small black rings whose horizon has the topology of  $S^1
\times S^{d-3}$ in a $d$-dimensional string theory. By analyzing the
singular black ring solution in the supergravity approximation, and
using various symmetries of the $\alpha'$ corrected effective action
we argue that the Bekenstein-Hawking-Wald entropy of the black
string solution in the full string theory agrees with the
statistical entropy of the same system up to an overall
normalization constant. While the normalization constant cannot be
determined by the symmetry principles alone, it can be related to a
similar normalization constant that appears in the expression for
small black holes without angular momentum in one less dimension.
Thus agreement between statistical and macroscopic entropy of
$(d-1)$-dimensional non-rotating elementary string states would
imply a similar agreement for a $d$-dimensional elementary string
state with spin. Our analysis also determines the structure of the
near horizon geometry and provides us with a geometric derivation of
the Regge bound. These studies give further evidence that a ring-like
horizon is formed when large angular momentum is added to a small
black hole.

\end{titlepage}


\section{Introduction and Summary}

Recently, there has been great deal of progress in computing corrections
to black hole entropy due to the effect of higher derivative terms in
string theory effective action and comparing the answer to the
statistical entropy of the same system \cite{Maldacena:1997de,
Behrndt:1998eq, LopesCardoso:1998wt, LopesCardoso:1999cv,
Mohaupt:2000mj, LopesCardoso:1999ur, LopesCardoso:2000qm,
Cardoso:2000fp, Dabholkar:2004yr, Dabholkar:2004dq, Sen:2004dp,
Hubeny:2004ji, Behrndt:2005he}.  A particularly interesting class of
examples is provided by the stringy `small' black holes. These are
singular solutions of the classical supergravity equations of motion
since they have vanishing area of the event horizon. However
microscopically they are described by BPS states of the fundamental
string and hence have non-zero degeneracies.  A simple class of such
examples is provided by the FP system or the so-called Dabholkar--Harvey
(DH) system \cite{Dabholkar:1989jt}, which is obtained by winding a
fundamental heterotic string $-w$ times around a circle $S^1$ and
putting $n$ units of momentum along the same circle. If the right-movers
are in the ground state then such a state is BPS but it can carry
arbitrary left-moving oscillations.\footnote{In the literature one often
finds two different sets of conventions. The first one uses the
convention that for a BPS state the momentum and winding along $S^1$
will have the same sign. The second one uses the convention that for a
BPS state the momentum and winding along $S^1$ will have opposite sign.
Here we are using the second convention.} Its microscopic entropy is
given by
\begin{eqnarray}
    S_{\rm micro}&=4\pi\sqrt{nw} \,. \label{S_micro_SBH}
\end{eqnarray}
Although in the supergravity approximation the corresponding solution
has zero horizon area and hence zero entropy, one expects that the
result will be modified by the higher derivative corrections since close
to the (singular) horizon the curvature and other field strengths become
strong and the supergravity approximation is expected to break
down. However one finds that the string coupling constant near the
horizon is small and hence one need not worry about string loop
corrections. By analyzing the behavior of the supergravity solution near
the horizon and using various symmetries of the full string tree level
effective action one finds that the $\alpha'$-corrected entropy must
have the form \be \label{eoldmac} C\sqrt{nw}\, , \ee for some constant
$C$ \cite{Sen:1995in, Peet:1995pe, Sen:1997is, Sen:2005kj}. However, the
constant $C$ cannot be calculated based on symmetry principles
alone. Recently for the special case of four dimensional small black
holes this constant was calculated in \cite{Dabholkar:2004yr} using a
special class of higher derivative terms in the effective action and was
found to give the value $4\pi$ in precise agreement with the microscopic
answer. Later it was demonstrated in \cite{Kraus:2005vz, Kraus:2005zm,
Kraus:2006wn} that other higher derivative corrections do not change the
result. Furthermore, it was shown that using special ensembles to define
entropy, the agreement between microscopic and macroscopic counting can
be pushed beyond the leading order to all orders in the large charge
expansion \cite{Dabholkar:2004yr, Sen:2005pu, Dabholkar:2005by,
Dabholkar:2005dt}.

The next natural question is what will happen if we add angular momentum
$J$ to the system.  Studying microscopic states and making use of 4D-5D
connection \cite{Bena:2004tk, Bena:2005ay, Gaiotto:2005gf,
Elvang:2005sa, Gaiotto:2005xt, Bena:2005ni}, it was argued in
\cite{Iizuka:2005uv} that if large angular momentum is added to a small
black hole in five dimensions with horizon topology $S^3$, it will turn
into a small black ring whose horizon has 
topology of $S^1 \times S^2$. The size of $S^2$ is 
of the order of the string scale, and hence is
small compared with the typical horizon scales of 
ordinary supersymmetric black ring solutions in 
\cite{Elvang:2004rt, Bena:2004de, Elvang:2004ds, Gauntlett:2004qy}
for which supergravity approximation is good.
Furthermore, in \cite{Iizuka:2005uv}, it was shown that the entropy of
small black rings can be related to that of four dimensional small black
holes,\footnote{Using the fuzzball picture \cite{Mathur:2005zp,
Mathur:2005ai} it was also argued in \cite{Lunin:2002qf} that this
system becomes a black ring.} and that the entropy calculation based on
this argument matches the independent entropy calculation of the
rotating DH system by Russo and Susskind \cite{Russo:1994ev} (see also
\cite{Palmer:2004gu, Marolf:2004fy, Bena:2004tk, Kraus:2005vz,
Bak:2004kz}).

Since a ring-like horizon is expected to show up for large $J$, it
is interesting to study the near horizon geometry of small black
rings from a macroscopic, geometric point of view. In particular we
would like to know the $J$ dependence of the horizon geometry.  One
would also like to generalize the results to higher dimensions.
These are the problems we shall address in this paper. In particular
we show that in arbitrary dimensions the supergravity equations of
motion admit a singular black ring solution carrying the same charge
quantum numbers as that of a rotating elementary string. This
solution is characterized by the charges $n$ and $w$ introduced
earlier, the angular momentum $J$ and a dipole charge $Q$ that
represents how many times the fundamental string winds in the
azimuthal directions. The microscopic entropy of such a system can
be easily computed from the spectrum of elementary string states
with spin and gives the answer 
\be\label{earbdim}
S_{stat}=4\pi\sqrt{nw-JQ}\, . 
\ee 
These states exist only for $nw\ge JQ$, which can be regarded as the
Regge bound for BPS states. On the other hand since the supergravity
solution is singular we cannot directly calculate the macroscopic
entropy. Nevertheless by examining the solution in a region of low
curvature, where supergravity approximation is expected to be valid, we
find that the space-time associated with the solution develops closed
time-like curves for $nw<JQ$ \cite{Emparan:2001ux}. 
Thus absence of such curves 
requires 
$nw>JQ$, thereby providing a geometric derivation of the Regge 
bound.\footnote{For oscillating 
string solutions describing string states with
angular momentum, a derivation of the Regge bound was given earlier in
\cite{Dabholkar:1995nc} by requiring regularity of the source terms in
the equation of motion.}  
Furthermore, by using a
symmetry argument very similar to that in
\cite{Sen:1995in,Peet:1995pe,Sen:1997is,Sen:2005kj} we show that the
entropy, if non-zero, is given by
\be \label{emacro} 
S_{BH}=C\sqrt{nw-JQ}\, , 
\ee 
for some constant $C$.  Furthermore $C$ is the same constant that would
appear in the expression \refb{eoldmac} for the macroscopic entropy of a
non-rotating black hole in one less dimension. Thus agreement between
microscopic and macroscopic entropy of a $(d-1)$-dimensional
non-rotating small black hole would imply a similar agreement for
$d$-dimensional rotating small black rings. This analysis also
determines the near horizon geometry in terms of some unknown constants
of order unity. In particular we can estimate the sizes of various
circles and spheres near the horizon.

The rest of the paper is organized as follows.  In section
\ref{smicro} we briefly review the computation of the statistical
entropy of rotating elementary string states leading to the answer
\refb{earbdim}. In section \ref{scaling} we describe the
supergravity solution describing the rotating small black ring and
study the effect of higher derivative corrections to this solution,
eventually arriving at the formula \refb{emacro} for the black ring
entropy. This analysis also determines the dependence of the near
horizon geometry on various charges in terms of some unknown
universal functions of the radial coordinate.  In section
\ref{sec:attractor/scaling} we use the entropy function formalism
\cite{Sen:2005wa} and scaling arguments for the case of ring-like
horizons, and show that the results are consistent with the ones
derived in section \ref{scaling}. This analysis determines the near
horizon geometry in terms of some unknown constants of order unity.
We conclude in section \ref{sec:disc} with a discussion of our
results and some open issues.

\medskip

\section{A Brief Review of the Microscopic Counting}
\label{smicro}

Let us consider
heterotic string theory in $\bbR_t\times \bbR^{d-1}\times
S^1\times T^{9-d}$ with $4\le d\le 9$ with $\bbR_t\times \bbR^{d-1}$
denoting $d$-dimensional Minkowski space. Our objects of interest
are the BPS elementary string excitations in this theory carrying
$n$ units of momentum and $-w$ units of winding charge
along the circle $S^1$ and angular momentum $J$ in a two
dimensional plane of $\bbR^{d-1}$. In the Ramond sector
such states are obtained
by acting with left-moving oscillators of level $N=nw+1$
on the Fock vacuum carrying charges $(n,w)$. In the NS sector
we also need to act on the Fock vacuum by a right-moving
fermionic oscillator of level 1/2. Let $x^1$ and $x^2$ denote
the coordinates of the two dimensional
plane in which the state carries angular momentum $J$.
Then if we define $x^\pm = (x^1\pm i x^2)/\sqrt 2$ and denote
by $\alpha^\pm_{-m}$ the corresponding oscillators of level $m$,
then the quantum number $J$ is given by the difference between
the number of
$\alpha^+_{-m}$ and the number of $\alpha^-_{-m}$ oscillators
acting on the
Fock vacuum. The statistical entropy is then given by the logarithm of
the number
of elementary string states subject to the restriction that the number of
$\alpha^+_{-m}$ oscillators minus the number of $\alpha^-_{-m}$
oscillators acting on the Fock vacuum is precisely $J$.

In our analysis we shall examine a finer definition of the statistical
entropy of rotating elementary strings
introduced in \cite{Dabholkar:2005qs}.
Let $Q$ be a positive integer. Let us further assume, for definiteness,
that $J$ is positive. We now count all states of the form
\be\label{eforma}
(\alpha^+_{-Q})^J \OO|vac\rangle
\ee
 where $\OO$ is an
arbitrary combination of left-moving oscillators of level $N_\OO
=nw-QJ+1$,
and $|vac\rangle$ denotes the
Fock vacuum for Ramond sector states, and the Fock vacuum acted
on by a right-moving fermionic oscillator of level 1/2 for the
NS sector states.
The statistical entropy of such states can be easily computed and
gives the answer \cite{Dabholkar:2005qs}
\be\label{eansstat}
S_{stat} \simeq 4\pi\sqrt{N_\OO}\simeq 4\pi \sqrt{nw - JQ}\,
\ee
for large $nw-JQ$.
Note that as defined, $\OO$ can have any spin and hence the states
which are counted by this procedure do not have definite
angular momentum. However
the dominant contribution to the entropy defined
this way comes from the states for which $\OO$ has spin close
to zero,
i.e.\
states  with spin close to
$J$. Thus even if we restrict $\OO$
to have total spin zero the result for entropy would
have the form given in \refb{eansstat}
at the leading order.

By restricting to states of the form
\refb{eforma} we have restricted to an exponentially
small subset of all
the states carrying angular momentum $J$. In particular the
dominant contribution to the statistical entropy for a fixed $J$
comes from states where most of the angular momentum is created
by the $\alpha^+_{-1}$ oscillators, i.e.\ from the states with $Q=1$.
The reason for considering states
of the form \refb{eforma} however is that we can distinguish states
with different $Q$ by the fields they produce even though $Q$ is
not a conserved charge. In particular the operator
$\alpha^+_{-Q}$ creates quanta of a mode for which the space-time
coordinates $x^1$ and $x^2$ have $Q$ complete oscillations as we
go once around the string. Thus a state of the form \refb{eforma}
where large number of these modes condense should correspond to
a configuration where the projection of the string in the $x^1$-$x^2$
plane
has $Q$ units of winding around its center of mass coordinate. As a
result
we expect that the corresponding gravity solution will be represented
by a ring like structure in the $x^1$-$x^2$ plane with $Q$ units of
winding charge along the ring. As we shall see,
the entropy of the corresponding solution will match the statistical
entropy given in eq.\ \refb{eansstat}.

\section{Scaling Analysis for Black Rings} \label{scaling}

In this section  we shall study the geometry of a small black ring
in arbitrary dimensions. Although in the supergravity
approximation this solution is singular along the ring, by
studying carefully the geometry near the singularity we shall
be able to determine the dependence of the entropy on various
charges {\it assuming that the higher derivative corrections modify
the solution in  a way such that it has
a finite entropy.} This will essentially involve generalization of the
scaling analysis
of \cite{Sen:1995in,Peet:1995pe,Sen:1997is,Sen:2005kj}
to the case of rotating black rings.

Consider heterotic string theory in $\bbR_t\times \bbR^{d-1}\times
S^1\times T^{9-d}$ with $4\le d\le 9$. Since in our analysis the
moduli associated with the torus $T^{9-d}$ will be frozen
completely, and furthermore all the mixed components of the metric
and the anti-symmetric tensor field with one leg along $T^{9-d}$ and
another leg along one of the other directions will be set to zero,
it will be more convenient to regard the theory as a theory in
$(d+1)$ space-time dimensions.\footnote{In other words we consider a
solution for which the world-sheet theory of the fundamental string
propagating in this background is a direct sum of two conformal
field theories. The first one is a free field theory associated with
the coordinates of $T^{9-d}$ and the sixteen additional left-moving
world-sheet bosons of the heterotic string theory. The second one is
an interacting theory associated with the string propagation in the
$(d+1)$-dimensional black ring solution. Our focus will be on the
latter theory.} The massless fields in $(d+1)$ space-time dimensions
that are relevant for analyzing the black ring solution are the
string metric $G_{\mu\nu}$, the anti-symmetric tensor field
$B_{\mu\nu}$ and dilaton $\Phi_{d+1}$. The string tree level action
involving these fields has the form \be\label{etree} \SSS = \int\,
d^{d+1}x \, \sqrt{-\det G} \, e^{-2\Phi_{d+1}}\, \LL\, , \ee where
$\LL$ is a function of the metric, Riemann tensor, the 3-form field
strength \be \label{edefh} H_{\mu\nu\rho}=\p_{\mu}B_{\nu\rho} +
\hbox{cyclic permutations of $\mu, \nu, \rho$} +
\Omega^{CS,L}_{\mu\nu\rho}(G)\, , \ee covariant derivatives of these
quantities, as well as covariant derivatives of the dilaton
$\Phi_{d+1}$ but not of $\Phi_{d+1}$ itself. Here
$\Omega^{CS,L}_{\mu\nu\rho}(G)$ denotes the Lorentz Chern-Simons
3-form constructed out of the string metric $G_{\mu\nu}$ -- more
precisely out of the spin connection compatible with this metric --
with some appropriate coefficient. Since we set all the gauge fields
in $(d+1)$-dimensions to zero, there are no gauge Chern-Simons term
in the definition of $H_{\mu\nu\rho}$. In the supergravity
approximation where we have only two derivative terms, the Lorentz
Chern-Simons term disappears from the expression for
$H_{\mu\nu\rho}$ and $\LL$ takes the form \be \label{esugra} \LL =
{1\over  (2\pi)^{d-2}(\alpha')^{(d-1)/2} }\, \left[R_G + 4
G^{\mu\nu} \p_\mu \Phi \p_\nu\Phi -{1\over 12} G^{\mu\mu'}
G^{\nu\nu'} G^{\rho\rho'} H_{\mu\nu\rho} H_{\mu'\nu'\rho'}\right]\,
. \ee

We denote the coordinates of
$\bbR_t$  and $S^1$  by $t$ and
$x^d$ respectively.
Let the coordinate
radius of $x^d$ direction be $R_d$.
For $\bbR^{d-1}$ we use a special coordinate system in which the
flat metric on $\bbR^{d-1}$ takes the form:
\begin{eqnarray}
 d\xv_{d-1}^2
 &=& {R^2\over (x-y)^2}
  \left[
   {dy^2\over y^2-1}+(y^2-1)d\psi^2+{dx^2\over
1-x^2}+(1-x^2)d\Omega_{d-4}^2
  \right],
\end{eqnarray}
where
$d\Omega_{d-4}$ denotes the line elements on the
$(d-4)$-sphere, $R$ is a constant whose value is given in
\refb{newrel_qn_gen}, and $x$, $y$ take values in the range
\be\label{erange}
-1\le x\le 1,  \qquad -\infty< y \le -1\, .
\ee
The relationship between these coordinates and the cartesian coordinates
of ${\bf R}^{d-1}$ has been given in
eqs.\ \refb{exyz1}, \refb{exyz2}.

A black ring solution in the supergravity approximation, describing a
rotating fundamental string of the type described in section
\ref{smicro}, has been constructed in Appendix \ref{sec:SBR_gen_dim}
based on \cite{Callan:1995hn, Dabholkar:1995nc, Lunin:2001fv,
Balasubramanian:2005qu} and takes the form:
\begin{eqnarray}
 ds^2_{str,d+1}&=&
  f_f^{-1}[-(dt-A_i dx^i)^2+(dx^d-A_i dx^i)^2+(f_p-1)(dt-dx^d)^2]
  + d\xv_{d-1}^2
  \nonumber\label{newmetric_SBR_gen}\\
 e^{2\Phi_{d+1}}&=& g^2 \, f_f^{-1},\qquad
 B_{td}= -(f_f^{-1}-1),\qquad
  B_{ti}=-B_{di}= f_f^{-1}\, A_i,
\end{eqnarray}
where $i=1,2,\dots,d-1$,
\begin{eqnarray}
 f_f &=& 1 + \frac{Q_f}{R^{d-3}} \left(\frac{x-y}{-2y}\right)^{(d-3)/ 2}
        {}_2 F_1\left(\frac{d-3}{4},\frac{d-1}{4}; 1; 1-\frac{1}{y^2}\right),
        \nonumber\\
 f_p &=& 1 + \frac{Q_p}{R^{d-3}} \left(\frac{x-y}{-2y}\right)^{(d-3)/ 2}
        {}_2 F_1\left(\frac{d-3}{4},\frac{d-1}{4}; 1; 1-\frac{1}{y^2}\right),
        \label{newfffpA_xy}\\
 A_i dx^i &=&-\frac{d-3}{2}\frac{q}{R^{d-5}}
            \frac{(y^2-1)(x-y)^{(d-5)/2}}{(-2y)^{(d-1)/2}}
        \,{}_2 F_1\left(\frac{d-1}{4},\frac{d+1}{4}; 2; 1-\frac{1}{y^2}\right)
        \, d\psi,
        \nonumber
\end{eqnarray}
and $Q_f$, $Q_p$, $q$ and $R$
are related to the quantized charges
$n$, $w$,  $Q$ and the angular momentum $J$ via the relations
\begin{equation}
 q={16\pi G_d\over (d-3)\Omega_{d-2}\alpha'}Q,\qquad
 R^2=\alpha'{J\over Q},\qquad
 Q_f={16\pi G_d R_d\over (d-3)\Omega_{d-2}\alpha'}w,
  \qquad
 Q_p={16\pi G_d\over (d-3)\Omega_{d-2}R_d}n\, .
  \label{newrel_qn_gen}
\end{equation}
Here $\Omega_D$ is the area of unit $D$-sphere
 and $G_d$ is the $d$-dimensional
Newton's constant obtained by regarding the
$S^1$ direction as compact
\begin{eqnarray} \label{enewton}
 16\pi G_d&=&{16\pi G_{d+1}\over 2\pi R_d}
  ={(2\pi)^{d-3} g^2\ap^{(d-1)/2}\over R_d }.
\end{eqnarray}
$n$ and $-w$ denote respectively
the number of units of momentum
and winding charge along the $S^1$ direction labeled by $x^d$.
$Q$ represents the number of units of winding charge along the
singular ring situated at $y=-\infty$. From the perspective of an
asymptotic observer $Q$ appears as a dipole charge and does not
represent a conserved gauge charge. In order that the metric
given in \refb{newmetric_SBR_gen} has the standard signature,
we require $R^2>0$, i.e.\
\be \label{ejq2}
JQ > 0\, .
\ee

For $d=5$ the hypergeometric functions simplify and we have
\begin{eqnarray}
 f_{f}&=&1+{Q_{f}(x-y)\over 2R^2},\qquad
 f_{p}=1+{Q_{p}(x-y)\over 2R^2},\qquad
 A_i dx^i=-{q\over 2}(1+y)d\psi.
\end{eqnarray}
This $d=5$ solution can be obtained by setting to zero one of the three
charges and two of the three dipoles charges of the supersymmetric black
ring solution of \cite{Elvang:2004rt, Bena:2004de, Elvang:2004ds,
Gauntlett:2004qy}.  
The solution with general $d$ can also be found by U-dualizing the
supergravity supertube solution of \cite{Emparan:2001ux}.

For reasons that will become clear later, we shall work with the
following assignment of charges:
\be \label{echargerange}
J \gg Q \gg 1, \qquad n \sim w, \qquad nw \sim J Q, \qquad
1-{JQ\over nw} \sim 1\, .
\ee
Since the ring singularity occurs as $y\to -\infty$ we shall
study the geometry near the singularity by examining the large
negative $y$ region. For
\be \label{eapproxrange}
{R\over \sqrt{\ap}} \gg |y| \gg
\left({1\over g^2 \, Q}\right)^{1\over d-4}
{R\over \sqrt{\ap}}\, ,
\left({R_d^2\over g^2 \, Q}\right)^{1\over d-4}
{R\over \sqrt{\ap}}\, ,1\; ,
\ee
the functions
$f_f$, $f_p$ and $A_\psi$ take the form:
\begin{eqnarray} \label{easymp}
f_f &\simeq&
 \left\{
  \begin{array}{ll}
   c_d   Q_f R^{3-d} |y|^{d-4} & \mbox{for $d>4$,}\\[1ex]
   \frac{1}{\pi} Q_f R^{-1}  \log|y| & \mbox{for $d=4$,}
  \end{array}
 \right.
\nonumber \\[1ex]
 f_p &\simeq&
  \left\{
  \begin{array}{ll}
   c_d   Q_p R^{3-d} |y|^{d-4} & \mbox{for $d>4$,} \\[1ex]
    \frac{1}{\pi} Q_p R^{-1}  \log|y|  & \mbox{for $d=4$,}
    \end{array}
 \right.
 \nonumber \\[1ex]
 A_\psi &\simeq&
  \left\{
   \begin{array}{ll}
  -   c_d   q R^{5-d} |y|^{d-4} & \mbox{for $d>4$,} \\[1ex]
   - \frac{1}{\pi} q R       \log|y| & \mbox{for $d=4$,}
   \end{array}
   \right.
\end{eqnarray}
where
\be\label{edefcd}
c_d = \frac{ \Gamma(\frac{d-4}{2}) }{ 2\sqrt{\pi}\,\Gamma(\frac{d-3}{2}) }\, .
\ee

Let us restrict to the case $d\ge 5$ and use the convention
$\alpha'=1$.
Using eqs.\ \refb{newrel_qn_gen} and \refb{enewton}
we see that in the region \refb{easymp} the original solution
\refb{newmetric_SBR_gen} takes the form:
\ben \label{einter}
ds_{str,d+1}^2 &=& {n\over w} {1\over R_d^2}\,
(dx^d-dt)^2 + 2 \, {(d-3)\Omega_{d-2}\over c_d \, (2\pi)^{d-3}}\
{1\over w}\, {R\over g^2}\,
\left(-{R\over y}\right)^{d-4}
\, dt\,  (dx^d-dt) \nonumber \\
&&
+ 2 {J\over w \, R_d}\, d\psi\, (dx^d-dt) +
 R^2 \, {d y^2\over y^4} +R^2 \, d\psi^2
+ \left(-{R\over y}\right)^2 d\Omega_{d-3}^2\,,
\nonumber \\
{1\over 2} B_{\mu\nu} dx^\mu \wedge dx^\nu &=&
-{(d-3)\Omega_{d-2}\over c_d \, (2\pi)^{d-3}}\,
{1\over w}\, {R\over g^2}\, \left(-{R\over y}\right)^{d-4}
\, dt\wedge (dx^d-dt) \nonumber \\ &&
+   {J\over w\, R_d} d(x^d-t) \wedge d\psi
+\hbox{constant}\,, \nonumber \\
e^{2 \Phi_{d+1}} &=&  {(d-3)\Omega_{d-2}\over c_d \,
(2\pi)^{d-3}}\,
{R\over w}
\,  \left(-{R\over y}\right)^{d-4} \, ,
\een
where
\be \label{defdw}
d\Omega_{d-3}^2
\equiv {dx^2\over 1-x^2} + (1-x^2) d\Omega_{d-4}^2
\ee
is the squared line element on a $(d-3)$-sphere and the constant in the
expression for $B$ represents a constant 2-form proportional to
$dt\wedge dx^d$ which can be removed by gauge transformation.  For this
solution all scalars constructed out of curvatures and other field
strengths are small in the region \refb{eapproxrange}.  For example the
$(d+1)$-dimensional Ricci scalar in the string frame goes as $(y/R)^2$
in this region.  As a result the supergravity approximation is still
valid in this region and the form of the solution \refb{einter} can be
trusted.

It is easy to see from \refb{einter} that
the $2\times 2$ matrix describing the metric in the
$\psi$-$x^d$ plane develops
a negative eigenvalue for $JQ > nw$.
Since the $\psi$-$x^d$ plane
is topologically a two dimensional torus,  the corresponding
space-time has closed time-like curves.
Thus absence of closed time-like
curve requires that
\be\label{ejq}
JQ \le nw\, .
\ee
Thus is precisely the Regge bound. The  fact that \refb{ejq}
can be derived by requiring absence of closed time-like curves
was noted in \cite{Emparan:2001ux} in a different U-duality
frame. Here we see that this geometrical condition 
is identical to the Regge bound
in the spectrum of elementary BPS string states that follows from the
left-right level matching condition in the microscopic
theory analyzed in section \ref{smicro}.

Note that both the conditions $JQ>0$ given in \refb{ejq2} and the Regge 
bound \refb{ejq} are
expected from the profile of the microscopic string 
underlying the solution as discussed in the appendix. This 
construction led to a 
manifestly positive $JQ$ and $(nw-JQ)$. The point however is that once we 
have obtained the solution, we can interpret it as a solution of the 
supergravity equations of motion parametrized by the charges $J$, $Q$, 
$n$ and $w$ without worrying about where it came from. 
Nevertheless regularity of the space-time geometry requires 
both 
the conditions \refb{ejq2} and \refb{ejq} to be satisfied.

{}From now on we shall restrict our analysis to the case $JQ<nw$.
By examining the solution \refb{einter} we see that this seems to
depend
on the various charges $n$, $w$, $J$, $Q$ as well as the asymptotic
values $g$ and $R_d$ of the moduli fields. We shall now show that
by making a suitable coordinate transformation the solution can be
made to be independent of
the parameters $g$ and $R_d$, and have simple dependence on
the charges.
We define
\ben \label{escaleco}
&& \sigma = \sqrt{{n\over w} -{JQ\over w^2}}
\, {1\over R_d}\, (x^d - t), \quad
\rho = -{R\over y}\, , \nonumber \\ &&
\tau = {(d-3)\Omega_{d-2}\over c_d \, (2\pi)^{d-3}}\,
{R_d  \over \sqrt{nw-JQ}} \, {R\over g^2} \, t,
\quad
\chi = \sqrt{J\over Q} \, \psi +{ \sqrt{JQ}\over w} \,
\, {1\over R_d}\, (x^d - t),
\een
In this coordinate system the region (\ref{eapproxrange})
gets mapped to
\be \label{enewappr}
1 \ll \rho \ll \left({g^2 Q}\right)^{1\over d-4},
\left({g^2 Q/R_d^2}\right)^{1\over d-4},R\; ,
\ee
and the coordinates $\chi$ and $\sigma$ have the following periods:
\be \label{eper2}
(\sigma,\chi) \equiv \left(\sigma, \chi+2\pi \sqrt{J\over Q}\right)
\equiv \left(\sigma+2\pi\sqrt{n\over w}\sqrt{1 -{JQ\over nw}},
\chi+ 2\pi {\sqrt{JQ}\over w}\right)\, .
\ee
The $\sigma$-$\chi$ plane at fixed values of the
other coordinates is topologically
a two dimensional torus of
coordinate area
\be\label{earea}
A_{\sigma\chi}= 4\pi^2 \sqrt{Jn\over Qw}
\sqrt{1-{JQ\over nw}}.\ee
In terms of the coordinates \refb{escaleco}
 the field configuration given in
\refb{einter} in the region
\refb{enewappr} takes the form:
\ben \label{escaledfield}
ds_{str,d+1}^2 &=& d\sigma^2
+ d\chi^2+ 2 \rho^{d-4} \, d\tau d\sigma
+ d\rho^2
+\rho^2 d\Omega_{d-3}^2 \nonumber \\
{1\over 2} B_{\mu\nu} dx^\mu \wedge dx^\nu &=&
-\rho^{d-4} \, d\tau\wedge d\sigma + \left({nw\over JQ}-1\right)^{-1/2}
d\sigma \wedge d\chi
\nonumber \\
e^{2 \Phi_{d+1}} &=& {(d-3)\Omega_{d-2}\over c_d \,
(2\pi)^{d-3}}\,
{1\over w}\, \sqrt{J\over Q}
\, \rho^{d-4} \, .
\een
By examining the solution we see that in this coordinate system
the solution as well as the
periodicities of the $(\sigma,\chi)$ coordinates are independent
of the parameters $g$ and $R_d$. Furthermore the dependence
of the solution on the charges comes via some additive constants
in the expressions for $B_{\sigma\chi}$ and $\Phi_{d+1}$, and
in the periodicities of the $(\sigma,\chi)$ plane. We shall make
use of this observation later to determine how the $\alpha'$-corrected
solution depends on the charges.

We now note the following
properties of this background:
\begin{itemize}
\item For $\rho\gg1$ curvature and other field strengths associated with
      this configuration are small.  Hence the higher derivative
      corrections to the equations of motion are negligible. Since
      $J\sim nw/ Q\sim w^2/Q$ and $Q\gg 1$, the string coupling constant
      $e^{\Phi_{d+1}}$ is also small in this region showing that the
      string loop corrections are also negligible.
\item For $\rho\sim 1$ the curvature and other field strengths become of
      order unity and hence the $\alpha'$ corrections become important.
      However $e^{\Phi_{d+1}}$ continues to be small since for the
      choice of charges of the form \refb{echargerange}, ${1\over w}
      \sqrt{J\over Q}\sim {1\over Q}$ is small. Thus string loop
      corrections are not important.
\item If we naively put a `stretched horizon' at $\rho\sim 1$, and
      calculate the naive entropy from the area of the stretched
      horizon, spanned by $\chi$, $\sigma$ and the coordinates of
      $S^{d-3}$ we get an answer proportional to $\sqrt{nw - QJ}$. This
      seems to agree with the formula \refb{eansstat} for the
      statistical entropy. However we should keep in mind that at
      $\rho\sim 1$ higher derivative corrections are important and
      neither the form of the solution nor the Bekenstein-Hawking
      formula for the entropy can be trusted. This is the problem to
      which we shall now turn.
\end{itemize}

The analysis of higher derivative corrections to the
solution given in \refb{escaledfield} is facilitated by the following
observations:
\begin{enumerate}
\item The solution given in \refb{escaledfield} is independent of the
      coordinates $\tau$, $\sigma$ and $\chi$. It also has a $SO(d-2)$
      spherical symmetry acting on the coordinates of the unit
      $(d-3)$-sphere whose line element has been denoted by
      $d\Omega_{d-3}$. We expect that the $\alpha'$-corrected solution
      will also preserve these symmetries.
\item Since the string coupling constant at the horizon is small, we can
      ignore string loop corrections to the effective action.  The tree
      level $\alpha'$ corrected theory can be described by an effective
      Lagrangian density in $(d+1)$ dimensions which does not depend on
      the periods of the $\sigma$ and the $\chi$ coordinates.  Since we
      are looking for solutions which are independent of the $\sigma$
      and $\chi$ coordinates, the independence of the Lagrangian density
      on their periodicities guarantees that there is no dependence of
      the solution on the periodicities of these variables.\footnote{One
      could worry about possible corrections to the effective action due
      to world-sheet instantons wrapping the $\chi$-$\sigma$
      torus. However such contributions will be exponentially suppressed
      due to large area of this torus measured in the string metric.}
      In particular the solution will have precisely
      the same form even if the $\sigma$ and the $\chi$ coordinates had
      been non-compact.

\item By examining the form of the solution \refb{escaledfield} we see
      that except for an additive constant term $ \left({nw\over
      JQ}-1\right)^{-1/2} $ in the expression for $B_{\sigma\chi}$ and
      an additive term of the form ${1\over 2}\ln \left({1\over w}\,
      \sqrt{J\over Q}\right)$ in the expression for $\Phi_{d+1}$, the
      solution is independent of the various charges and the asymptotic
      values of the various moduli {\it e.g.\/}\ $g$, $R_d$ etc.  Since
      the tree level effective Lagrangian density depends on
      $\Phi_{d+1}$ only via an overall multiplicative factor of
      $e^{-2\Phi_{d+1}}$ and terms involving derivatives of $\Phi_{d+1}$
      and depends on $B_{\mu\nu}$ only through its field strength $dB$,
      $B_{\mu\nu}$ and $\Phi_{d+1}$ can be shifted by arbitrary
      constants without affecting the rest of the solution.  Thus we
      shall expect that even after including the $\alpha'$ corrections
      the solution continues to be independent of the various charges
      and the parameters $g$, $R_d$ etc.\ except for an additive factor
      of ${1\over 2}\ln \left({1\over w}\, \sqrt{J\over Q}\right)$ in
      $\Phi_{d+1}$ and an additive factor of $ \left({nw\over
      JQ}-1\right)^{-1/2} $ in $B_{\sigma\chi}$.
\end{enumerate}
The general form of the modified solution subject
to these requirements is given by
\ben \label{emodified}
ds_{str,d+1}^2 &=& g_{\alpha\beta}\left(\rho\right)
d\zeta^\alpha d\zeta^\beta + f_1\left(\rho \right) d\Omega_{d-3}^2
+d\rho^2\nonumber \\
{1\over 2} B_{\mu\nu} dx^\mu \wedge dx^\nu &=&
b_{\alpha\beta}\left(\rho
\right) d\zeta^\alpha\wedge d\zeta^\beta + \left({nw\over JQ}-
1\right)^{-1/2} d\sigma\wedge d\chi, \nonumber \\
e^{2\Phi_{d+1}} &=&
{1\over w}\, \sqrt{J\over Q} \,
f_2\left(\rho \right) \, ,
\een
where $\zeta\equiv(\zeta^0,\zeta^1,\zeta^2)$ stands collectively for the
coordinates $\tau$, $\sigma$ and $\chi$, and $g_{\alpha\beta}$,
$b_{\alpha\beta}$, $f_1$ and $f_2$ are some universal functions of the
coordinate $\rho$, independent of any other charges and parameters.  The
solution in the original coordinate system, if needed, can now be found
by applying the inverse of the coordinate transformation \refb{escaleco}
on \refb{emodified}.

{\it A priori} we do not know the form of the functions
$g_{\alpha\beta}$, $b_{\alpha\beta}$, $f_1$ and $f_2$, but let us
proceed with the assumption that $\alpha'$ corrections modify
the near horizon geometry to that of an (extremal) black hole.
In that case computation of the entropy requires us to
integrate certain combinations of the fields over the
horizon \cite{Wald:1993nt, Jacobson:1993vj,
Iyer:1994ys, Jacobson:1994qe}.
{}From the coordinate area of
$4\pi^2 \sqrt{Jn\over Qw}
\sqrt{1-{JQ\over nw}}$ in the $\chi$-$\sigma$ plane
we get a factor proportional to
$\sqrt{Jn\over Qw}
\sqrt{1-{JQ\over nw}}$
from integration along these coordinates. The multiplicative
factor of ${w}\, \sqrt{Q\over J}$ in
$e^{-2\Phi_{d+1}}$ appears as an overall normalization factor
in the $\alpha'$ corrected effective action and gives a factor
proportional to ${w}\, \sqrt{Q\over J}$
in the entropy. Besides these multiplicative factors the contribution
to the entropy cannot depend on any other charges or parameters
since  the solution has no
non-trivial dependence on any other parameter. This gives
\be \label{ebhen}
S_{BH} = C \sqrt{nw - JQ} \,
\ee
for some constant $C$. This is in precise agreement with the answer
\refb{eansstat} for the statistical entropy if we take $C=4\pi$.

It is worth emphasizing the role of the scaling region
\refb{enewappr} where the supergravity solution \refb{escaledfield}
is valid. As we have seen, in this region the dependence of the
solution on the asymptotic moduli parameters $R_d$ and $g$
disappears completely. This is then used to argue that the $\alpha'$
corrected solutions near the horizon will also be independent of
these parameters. Thus this scaling region acts as a shield which
isolates the near horizon region from the asymptotic region.
However, since for large but finite charges the scaling region has a
large but finite size, we expect that this shielding will work only
in the leading order, and could break down at the subleading order
in an expansion in inverse powers of charges. In section
\ref{sec:attractor/scaling} we shall see that if we assume that the
near horizon geometry has an $AdS_2$ factor then the independence of
the entropy of the asymptotic parameters holds to all orders since
there is an infinite throat region of $AdS_2$ that separates the
near horizon geometry from the asymptotic geometry by an infinite
amount.

We note in passing that if we had tried to carry out a similar scaling
analysis for the rotating black hole solutions of 
\cite{Sen:1994eb} or their generalization to higher dimension, we would be 
led to 
the conclusion that the result for the black hole entropy is of the form
$\sqrt{nw} \, g(J^2/nw)$ for some function $g$
\cite{unpublished}. 
This is in contradiction 
with the microscopic result which gives the entropy to be $\sqrt{nw}$ 
times a function of $J/nw$. This also shows that the 
ring geometry is the correct geometry for describing an
elementary 
string state  with spin.

Just based on the scaling analysis we cannot determine the
value of this coefficient $C$.
However since
from the point of view of the near horizon geometry the coordinate
$\psi$ describes a non-contractible circle, we can regard this as a
compact direction. In that case the solution
\refb{escaledfield} and its
$\alpha'$ corrected version \refb{emodified} can be
regarded as one describing the near horizon geometry
of a non-rotating black hole in a space-time
with $(d-1)$ non-compact dimensions,
carrying $n$ units of momentum and
$-w$ units of winding along the
$x^d$ directions and $J$ units of momentum and $Q$ units of
winding along the $\psi$ directions. Thus as long as non-rotating
small black holes in $(d-1)$ dimensions have finite entropy, rotating
$d$ dimensional black holes also have finite entropy. Furthermore
if the entropy of non-rotating small
black holes in $(d-1)$ dimensions agrees
with the microscopic entropy, the constant $C$ is equal to $4\pi$.
This in turn will imply that the entropy of the rotating small black
rings in $d$ dimension also agrees with the
corresponding statistical entropy.

Regarding the solution as a $(d-1)$-dimensional small black hole
also gives us some insight into the form of eq.\ \refb{ebhen}.
The $(d-1)$-dimensional effective action is known to have a
continuous
$SO(2,2)$ symmetry to all orders in $\alpha'$ expansion due to
the fact that we are examining a sector where the fields are independent
of $\sigma$ and $\chi$ directions. The only $SO(2,2)$ invariant
combination of the charges $n$, $w$, $J$ and $Q$
is
$nw-JQ$ or a function of this combination. Since \refb{ebhen}
depends on this combination we see that this formula is consistent
with the SO(2,2) invariance of the theory.

\section{Entropy Function and Near Horizon Geometry}
\label{sec:attractor/scaling}

In section \ref{scaling} we determined the geometry of the small
black ring in terms of some unknown universal functions
$g_{\alpha\beta}(\rho)$, $b_{\alpha\beta}(\rho)$, $f_1(\rho)$ and
$f_2(\rho)$. Supergravity approximation to the effective action,
which is valid for large $\rho$, determines the behavior of these
functions at large $\rho$. In this section we shall describe a
general procedure for determining the form of these functions at
small $\rho$, assuming that the near horizon geometry in this region
approaches that of an extremal black hole with an $AdS_2$ factor,
and as a result possesses an enhanced  isometry $SO(2,1)$. Our main
tool in this analysis will be the entropy function method described
in \cite{Sen:2005wa,Astefanesei:2006dd}.

As in section \ref{scaling} we consider heterotic string theory
compactified on $T^{9-d}\times S^1$ and consider an extremal black
ring solution in this theory. From the analysis of section
\ref{scaling} we know that the geometry close to the horizon has two
compact directions labeled by the angular coordinate $\psi$ and the
coordinate along $S^1$: \be \label{edefyd} y^d = (x^d-t)/R_d\, , \ee
each with period $2\pi$. Thus we can analyze the near horizon
geometry of such a black ring by analyzing the $(d-1)$-dimensional
theory obtained via dimensional reduction of the original theory on
these two circles. We parametrize these $(d-1)$ dimensions by the
coordinates $\{\xi^m\}$ with $0\le m\le (d-2)$, and introduce the
following $(d-1)$-dimensional fields in terms of the original
$(d+1)$-dimensional fields:\footnote{Our definition of the
$(d-1)$-dimensional antisymmetric tensor field $\wh B_{mn}$ differs
from the standard one, {\it e.g.\/}\ the one used in
\cite{Sen:2004dp}, by a term proportional to
$(A^{(1)}_{[m}A^{(3)}_{n]} + A^{(2)}_{[m}A^{(4)}_{n]})$. As a
consequence the expression for the gauge Chern-Simons term appearing
in the expression for $\wh H_{mnp}$  is also slightly different.}
\ben \label{edefd-1} ds_{str,d+1}^2 &=& \wh G_{mn}(\xi) d\xi^m
d\xi^n + R(\xi)^2 (dy^d + A_m^{(1)}(\xi)d\xi^m)^2 +
\wt R(\xi)^2 (d\psi + A_m^{(2)}(\xi)d\xi^m)^2 \nonumber \\
&& +
2\, S(\xi) \, (dy^d + A_m^{(1)}(\xi)d\xi^m)\,
(d\psi + A_n^{(2)}(\xi)d\xi^n)\, , \nonumber \\
{1\over 2}\, B_{\mu\nu} dx^\mu\wedge dx^\nu &=&
{1\over 2} \, \wh B_{mn}(\xi) \, d\xi^m \wedge d\xi^n
+ C(\xi) \, (dy^d + A_m^{(1)}(\xi)d\xi^m)\wedge
(d\psi + A_m^{(2)}(\xi)d\xi^m)
\nonumber \\
&& + (dy^d + A_m^{(1)}(\xi)d\xi^m)\wedge A^{(3)}_n(\xi) \, d\xi^n
+ (d\psi + A_m^{(2)}(\xi)d\xi^m)\wedge A^{(4)}_n(\xi)
\, d\xi^n\, . \nonumber \\
\een
Thus the fields in $(d-1)$ dimensions include a metric
$\wh G_{mn}$, an anti-symmetric tensor field $\wh B_{mn}$,
four gauge fields $A_m^{(i)}$ for $1\le i\le 4$ and five scalar
fields $R$, $\wt R$, $S$, $C$ and $\Phi_{d+1}$. The gauge invariant
field strengths constructed out of the fields $A_m^{(i)}$ and
$\wh B_{mn}$ are:
\ben \label{eadd2}
F^{(i)}_{mn} &=& \p_m A^{(i)}_n -
\p_n A^{(i)}_m, \quad i=1,2,3,4
\nonumber \\
\wh H_{mnp} &=&
\left(\p_m  \wh B_{np}  + \hbox{cyclic permutations of $m,n,p$}
\right)  + \Omega^{CS}_{mnp}\, ,
\een
where
\be \label{edefcsterm}
\Omega^{CS}_{mnp} = \left\{ (A^{(3)}_m F^{(1)}_{np}
+ A^{(4)}_m F^{(2)}_{np}) + \hbox{cyclic permutations
of $m,n,p$}\right\} + \Omega^{CS,L}_{mnp}(\wh G)
\, .
\ee
The action of the dimensionally reduced theory has the form:
\be \label{edaction}
\SSS=\int d^{d-1} x\, \sqrt{-\det \wh G}\, e^{-2\Phi_{d+1}}
\, \LL_{d-1}\, ,
\ee
where $\LL_{d-1}$ is a scalar function of the scalars $R$, $\wt R$,
$S$ and $C$,
the metric $\wh G_{mn}$, Riemann tensor, the field strengths
$F^{(i)}_{mn}$ and $\wh H_{mnp}$, $\p_m \Phi_{d+1}$
and covariant derivatives
of these quantities.

The presence of the Chern-Simons terms in the definition of $\wh
H_{mnp}$ makes this form of the action unsuitable for applying the
entropy function formalism since the latter requires the Lagrangian
density, when expressed in terms of the independent fields of the
theory, to be a function of manifestly covariant quantities like the
metric, Riemann tensor, gauge field strengths, scalar fields, covariant
derivatives of these quantities etc., and not {\it e.g.\/}\ of
non-covariant quantities like the gauge fields or the spin
connection.\footnote{The only exception to this rule are $p$-form gauge
fields whose near horizon values themselves are manifestly invariant
under all the isometries of the near horizon geometry. In this case we
never need to explicitly use the gauge invariance associated with these
fields and can regard them as ordinary tensor fields. \label{fo1}} To
get around this problem we need to dualize the theory
\cite{Sahoo:2006pm}.  For this note that $\wh H_{mnp}$ satisfies the
Bianchi identity
\be\label{ebianchi}
\p_{[m} \left(\wh H_{npq]} -
\Omega^{CS}_{npq]} \right) =0\, .
\ee
We now introduce a new $(d-5)$-form field $\BB_{m_1\ldots m_{d-5}}$,
define
\be \label{eadd5}
\HH_{m_1\ldots m_{d-4}} = \p_{m_1}
\BB_{m_2\ldots m_{d-4}} + \hbox{cyclic permutations of
$m_1\ldots m_{d-4}$ with sign}\, ,
\ee
to be its field strength and consider a new action
\be \label{enewac}
\int d^{d-1} x \, \sqrt{-\det \wh G}\,  \wt \LL_{d-1}\, ,
\ee
\be \label{enewlagold}
\sqrt{-\det  \wh G}\,  \wt \LL_{d-1}
= \left[\sqrt{-\det \wh G}\, e^{-2\Phi_{d+1}}\, \LL_{d-1}
+ \epsilon^{m_1\ldots m_{d-1}}\, \left(\wh H_{m_1m_2 m_3}
-\Omega^{CS}_{m_1m_2m_3}\right) \,
\HH_{m_4\ldots m_{d-1}}\right]\, ,
\ee
regarding $\wh H_{mnp}$ and
$\BB_{m_1\ldots m_{d-5}}$ as independent fields.
Here $\epsilon^{m_1\ldots m_{d-1}}$ is totally antisymmetric
in its indices, with $\epsilon^{01\ldots (d-2)}=1$.
Equations of motion of the $\BB$ field gives the Bianchi
identity \refb{ebianchi}. On the other hand the equations of motion
of $\wh H_{m_1m_2m_3}$ gives
\be \label{eeqh}
{\delta \SSS\over \delta \wh H_{m_1 m_2 m_3}}
+ \epsilon^{m_1\ldots m_{d-1}}\, \HH_{m_4\ldots m_{d-1}}=0\, .
\ee
Together with the Bianchi identity
$\p_{[m_3}\HH_{m_4\ldots m_{d-1}]}=0$,  \refb{eeqh}
gives us the original equations of
motion of the $\wh B_{mn}$ field:
\be \label{eeqhor}
\p_p\left( {\delta \SSS\over
\delta \wh H_{mnp}}\right)=0\, .
\ee
Thus classically \refb{enewac}
and \refb{edaction} gives rise to the same theory and we
can choose to work with the action \refb{enewac}.

The action \refb{enewac} can now be brought into
a manifestly gauge, local Lorentz and general coordinate
invariant form by integrating the last term in
\refb{enewlagold} by parts. This gives a new Lagrangian density:
\ben \label{enewlag}
\sqrt{-\det  \wh G}\,  \wt \LL_{d-1}'
&=& \sqrt{-\det \wh G}\, e^{-2\Phi_{d+1}}\, \LL_{d-1} \nonumber \\
&&
- (d-4) \epsilon^{m_1\ldots m_{d-1}}\, \p_{m_4}\,
 \left(\wh H_{m_1m_2 m_3}
-\Omega^{CS}_{m_1m_2m_3}\right) \,
\BB_{m_5\ldots m_{d-1}}\, .
\een
Since
$d\Omega^{CS}$ can be expressed as a function of manifestly
covariant quantities like the Riemann tensor and gauge field
strengths $F^{(i)}_{mn}$,  the
Lagrangian density $\wt\LL_{d-1}'$
is suitable for
applying the entropy function formalism.
Note however that $\wt\LL_{d-1}'$
does not have manifest  symmetry under the gauge transformation
$\BB\to \BB + d\Lambda$, $\Lambda$ being
a $(d-6)$-form gauge transformation
parameter.
This will not
affect our analysis since we shall be considering field configurations
for which the field $\BB$ (and not just its field strength $\HH$) has
all the necessary symmetries. Thus, as discussed in
footnote \ref{fo1}, we never need to make use of
the gauge invariance associated with the field $\BB$, and shall treat
$\BB$ as an independent tensor field.

We are now ready to apply the entropy function formalism. We begin
with the basic postulate that in terms of the $(d-1)$-dimensional fields
the near horizon metric of the extremal black ring has the structure
of $AdS_2\times S^{d-3}$, with all other field configurations
respecting the $SO(2,1)\times SO(d-2)$ isometry of
$AdS_2\times S^{d-3}$.  Then the
general near horizon geometry of the black ring is
of the form:\footnote{$r$
is related to the coordinate $\rho$ of section \ref{scaling} by
the relation $ r = e^{\rho/\sqrt v_1}$.
The coordinate $\bar t$ and $\tau$ are in general
related by a scaling. However since a rescaling of the form
$\bar t\to \lambda \, \bar t$, $r\to r/\lambda$, being an isometry
of $AdS_2$, preserves the form of
the solution, we can use this freedom to choose
$$
\bar t = \tau, \quad r = c\, e^{\rho/\sqrt v_1}\, ,
$$
for some constant $c$.
}
\ben\label{ne1.1}
&& \wh G_{mn} d\xi^m d\xi^n = v_1\,
\left(-r^2 d \bar t^2+{dr^2\over
r^2}\right) + v_2 \, d\Omega_{d-3}^2 \, ,
\nonumber \\
&&
R =u_{R}, \quad \wt R = \wt u_R, \quad S = u_S\, , \quad C=u_C, \quad
\Phi_{d+1}=u_\Phi \, , \nonumber \\
&& {1\over 2}\, F^{(i)}_{mn}dx^m\wedge dx^n
= e_i\, dr \wedge d \bar t
\, , \quad i=1,2,3,4 \nonumber \\
&& {1\over (d-5)!} \, \BB_{m_1\ldots m_{d-5}} \, d\xi^{m_1}
\wedge \ldots \wedge d\xi^{m_{d-5}}
= \cases{\hbox{$b$ for $d=5$}\cr \hbox{$b \, dr\wedge dt$
for $d=7$}
\cr
\hbox{0 for $d\ne 5,7$}}\, ,
\een
where $v_1$, $v_2$, $u_R$, $\wt u_R$,  $u_S$,  $u_C$,
$u_\Phi$,
$e_1$, $e_2$, $e_3$, $e_4$ and $b$ are
constants and $d\Omega_{d-3}$ denotes the line element on a
unit $(d-3)$-sphere. Note that we have not explicitly given the
$\wh H$ field background; we are implicitly assuming that $\wh H$
has been eliminated from the action using its equation of motion.
In any case, the only possible non-zero component of $\wh H_{mnp}$
consistent with the symmetries of $AdS_2\times S^{d-3}$ is a flux
through $S^{d-3}$ in the special case of $d=6$. However since we are
considering solutions without magnetic charge,
--- more specifically NS 5-brane charge --- even for $d=6$ this flux
should vanish. Thus we can consistently set $\wh H_{mnp}$ to zero.
The
entropy function for a black ring
carrying $n$ units of momentum and $-w$ units of winding
along $y^d$ and  $J$ units of momentum and $Q$ units of winding
along $\psi$, is now given
by \cite{Sen:2005wa}\footnote{According to the convention of
section \ref{scaling}, $n$, $J$, $w$ and $-Q$
represent the charges associated with
the $A^{(1)}_m$, $A^{(2)}_m$,
$A^{(3)}_m$ and $A^{(4)}_m$ fields respectively. This explains the
signs in front of various charges in \refb{ne1.2}. Note however that
we have chosen to call $-w$ and $Q$ the winding charges along
$y^d$ and $\psi$ respectively.}
\ben\label{ne1.2}
&&\EE(n, J, w, Q, v_1, v_2, u_R, \wt u_R,  u_S, u_C,
u_\Phi,
e_1,e_2,  e_3, e_4, b) \nonumber \\
 &=& 2\pi \left(n e_1 + J e_2  + w e_3 - Q e_4 -
\int_{S^{d-3}}\, \sqrt{-\det \wh G} \,
\wt \LL_{d-1}'\right)\, ,
\een
where  $\sqrt{-\det \wh G} \,
\wt \LL_{d-1}'$ has to be evaluated on the horizon.
The entropy is
given by $\EE$ after extremizing it with respect to the near horizon
parameters  $v_1$, $v_2$, $u_R$, $\wt u_R$,  $u_S$, $u_C$,
$u_\Phi$,
$e_1$, $e_2$, $e_3$, $e_4$ and $b$.

This gives a general algebraic procedure for determining the near
horizon geometry of a small black ring for a given action.  We shall now
show that the entropy calculated from this formalism has the same
dependence on $n$, $w$, $J$ and $Q$ as was derived in the previous
section.  For this we need to use the known scaling properties of the
$\alpha'$ corrected tree level effective action of the heterotic string
theory.  First of all note from \refb{enewlag} that
\be\label{ne2} \wt\LL_{d-1}' \to \lambda \, \wt\LL_{d-1}' \quad
\hbox{under} \quad e^{-2\Phi_{d+1}}\to\lambda \, e^{-2\Phi_{d+1}},
\quad \BB_{m_1\ldots m_{d-5}}\to \lambda \, \BB_{m_1\ldots
m_{d-5}}\, . \ee The freedom of changing the periodicity along the
circle $S^1$ labeled by $y^d$ and subsequently making a rescaling of
the $y^d$ coordinate to bring the period back to $2\pi$ gives
another scaling property of the Lagrangian density: \be \label{ne3}
\wt\LL_{d-1}' \to   \kappa\, \wt\LL_{d-1}' \quad \hbox{under} \quad
A^{(1)}_{m} \to \kappa^{-1} \, A^{(1)}_{m},
 \quad A^{(3)}_m\to \kappa\, A^{(3)}_m, \quad
R \to \kappa  \, R, \quad S\to \kappa \, S, \quad C\to \kappa \, C\, .
\ee
Finally
there is a similar scaling property associated with the
scaling of the $\psi$ coordinate. This gives
\be\label{ne10}
\wt\LL_{d-1}' \to
\eta\,
\wt\LL_{d-1}'
\quad \hbox{under}
\quad
A^{(2)}_m\to \eta^{-1} A^{(2)}_m, \quad
A^{(4)}_m\to \eta  A^{(4)}_m, \quad
\wt R\to \eta \, \wt R, \quad
S \to \eta  \, S, \quad C\to \eta\, C \, .
\ee
{}From \refb{ne2}, \refb{ne3} and \refb{ne10} we can derive the
following properties of the entropy function:
\be\label{ne4}
\EE \to \lambda\, \EE \quad \hbox{under} \quad
e^{-2u_\Phi}\to\lambda \, e^{-2u_\Phi}, \quad b\to\lambda\, b,
\quad Q\to \lambda \, Q, \quad
n\to \lambda \, n, \quad w\to \lambda \, w,
\quad J\to \lambda \, J\, ,
\ee
\ben \label{ne5}
\EE \to \kappa \EE  &\hbox{under}& \quad
e_1 \to \kappa^{-1} \, e_1, \quad n\to \kappa^2 \, n,
\quad
J\to \kappa J, \quad
e_3\to \kappa e_3,  \quad Q\to \kappa Q,  \nonumber \\
&& \quad u_R \to \kappa \, u_R, \quad
u_S\to \kappa \, u_S, \quad u_C\to \kappa \, u_C\, ,
\een
and
\ben \label{ne11}
\EE\to \eta\, \EE &\hbox{under}& \quad
n\to \eta\, n, \quad e_2\to \eta^{-1}\, e_2, \quad J\to \eta^2\, J, \quad
w\to \eta \, w, \quad  e_4 \to \eta \, e_4, \nonumber \\ &&
\wt u_R \to \eta\, u_R, \quad u_S\to \eta \, u_S, \quad u_C\to \eta \,
u_C\, .
\een
{}From this it follows that after elimination of the various
near horizon parameters by extremizing $\EE$,
the entropy $S_{BH}=\EE$
has the property:
\be\label{ne6}
S_{BH}\to \lambda \, S_{BH}
\quad \hbox{under} \quad
n\to \lambda \, n, \quad w\to \lambda \, w, \quad J\to \lambda \, J
\quad Q\to \lambda \, Q\, ,
\ee
\be\label{ne7}
S_{BH}\to \kappa\, S_{BH} \quad \hbox{under} \quad
n \to \kappa^2\, n, \quad
Q\to \kappa\, Q, \quad J\to \kappa\, J\, ,
\ee
and
\be\label{ne12}
S_{BH}\to \eta\, S_{BH} \quad \hbox{under} \quad
n\to \eta \, n, \quad w\to \eta \, w,
\quad J\to \eta^2 \, J\, .
\ee
Eqs.\ \refb{ne6}, \refb{ne7} and \refb{ne12} give
\be\label{ne8b}
S_{BH} = \sqrt{nw}\, f\left({JQ\over nw}\right)\, ,
\ee
for some function $f$.

We can constrain the form of the function $f$ by noting that
the dimensionally reduced Lagrangian density has a further symmetry
induced by a  rotation in the $y^d-\psi$ plane by the matrix
\be \label{edefu}
U = \pmatrix{\cos\theta & \sin\theta\cr -\sin\theta & \cos\theta}
\, .
\ee
This induces a transformation
\be \label{erotate}
\pmatrix{R^2 & S\cr
S & R^2} \to U \pmatrix{R^2 & S\cr
S & \wt R^2} U^T, \quad \pmatrix{A^{(1)}_m\cr A^{(2)}_m}
\to U \, \pmatrix{A^{(1)}_m\cr A^{(2)}_m}
, \quad \pmatrix{A^{(3)}_m\cr A^{(4)}_m}
\to U \, \pmatrix{A^{(3)}_m\cr A^{(4)}_m}\, .
\ee
Thus the entropy function $\EE$ is invariant
under
\ben \label{erotpar}
&& \pmatrix{e_1\cr e_2}
\to U \pmatrix{e_1\cr e_2}, \quad
\pmatrix{e_3\cr e_4}
\to U \pmatrix{e_3\cr e_4}, \quad
\pmatrix{n\cr J} \to U \pmatrix{n\cr J}, \quad
\pmatrix{w\cr -Q}\to U \pmatrix{w\cr -Q},  \nonumber \\
&&
\quad \pmatrix{u_R^2 & u_S\cr
u_S & \wt u_R^2} \to U \pmatrix{u_R^2 & u_S\cr
u_S & \wt u_R^2} U^T\, .
\een
As a result the black hole entropy $S_{BH}$ is invariant under
\be \label{erotsbh}
\pmatrix{w\cr -Q}\to U \pmatrix{w\cr -Q}, \quad
\pmatrix{n\cr J} \to U \pmatrix{n\cr J}\, .
\ee
Together with \refb{ne8b} this uniquely fixes the form of $S_{BH}$
to be
\be \label{ebhfinform}
S_{BH} = C \, \sqrt{nw - JQ}\, ,
\ee
for some constant $C$.

If the entropy function had no flat directions then we would also be
able to determine the near horizon parameters labeling the solution
by demanding that the solution remains invariant under the symmetry
transformations described above. This however is not possible due to
the existence of two flat directions of the entropy
function.\footnote{If we consider the case where the coordinates
$y^d$ and $\psi$ are both compact and there are no fluxes then the
vacuum moduli space, ignoring the $T^{9-d}$ factor, is
$SO(2,2)/(SO(2)\times SO(2))$ where $SO(2,2)$ represents the
continuous T-duality symmetry associated with a $T^2$
compactification, and $SO(2)\times SO(2)$ is its maximal compact
subgroup. Since the charge vector $(n,J,w,Q)$ is invariant under an
$SO(2,1)$ subgroup of this $SO(2,2)$ group, once we switch on a flux
proportional to this charge vector the moduli space of solutions
becomes the two dimensional space $SO(2,1)/SO(2)$. This would
correspond to two flat directions of the entropy function. The only
exceptions  are light-like charge vectors in which case the little
group will be SO(1,1), but this corresponds to the case $nw-JQ=0$
which we are not considering here. In general a symmetry
transformation of the type given in eqs.\ \refb{ne2}-\refb{ne10} and
\refb{erotate}, instead of leaving the solution unchanged, will take
one solution to another solution.} Thus for a given set of charges
$(n,J,w,Q)$ the entropy function extremization condition gives a two
parameter family of solutions. Which of these solutions actually
appear as the near horizon configuration of the black ring will
depend on the asymptotic data --- in particular the asymptotic
values of the moduli fields and the information that we are
considering a black ring in $d$-dimensions rather than a black hole
in $(d-1)$ dimensions where the $\psi$ direction remains compact
even asymptotically. Thus the entropy function itself cannot give
complete information about the near horizon geometry even if we knew
the full $\alpha'$ corrected action.

On the other hand, the analysis in section \ref{scaling} does know about
the asymptotic infinity since it is based on a solution of the
supergravity equations of motion embedded in an asymptotically flat
$d$-dimensional spacetime.  Therefore the solution \refb{escaledfield}
in the scaling region \refb{enewappr} does not involve any unknown
parameter, and in turn completely determines the general form
\refb{emodified} of the solution near the horizon.  This suggests that
we can combine the results of section \ref{scaling} and this section to
fix the form of the near horizon geometry. We first note that the
solution \refb{ne1.1} corresponds to the $(d+1)$-dimensional field
configuration:
\ben\label{ed+1dim}
ds_{str,d+1}^2 &=& v_1\,
\left(-r^2 d \bar t^2+{dr^2\over
r^2}\right) + v_2 \, d\Omega_{d-3}^2
+ u_R^2 (dy^d + e_1 r d\bar t)^2 + \wt u_R^2 (d\psi+e_2 r\, d\bar t)^2
\nonumber \\ &&
+ 2\, u_S\, (dy^d + e_1 r d\bar t)\, (d\psi+e_2 r d\bar t)\, , \nonumber \\
{1\over 2} B_{\mu\nu} dx^\mu\wedge dx^\nu &=&
u_C \, (dy^d + e_1 r d\bar t)\wedge (d\psi+e_2 r d\bar t)
+ e_3 r \, (dy^d + e_1 r d\bar t) \wedge d\bar t \nonumber \\
&& + e_4 r \,
(d\psi+e_2 r d\bar t)\wedge d\bar t\, ,
\nonumber \\
e^{-2\Phi_{d+1}} &=& e^{-2u_\Phi}\, .
\een
In writing \refb{ed+1dim} we have used the coordinate transformations 
$y^d\to y^d + a\bar t$, $\chi\to \chi+b\bar t$ to remove the constant 
terms in $A^{(1)}_t$ and $A^{(2)}_t$. 
On the other hand \refb{escaleco},
\refb{emodified} tells us
that if we use the $(\sigma,\chi)$ coordinate
system defined through
\be\label{esigmachi}
\sigma=\sqrt{{n\over w} - {JQ\over w^2}}\, y^d, \qquad
\chi = \sqrt{J\over Q} \, w + {\sqrt{JQ}\over w}\, y^d\, ,
\ee
then the solution has a universal form except for 
some
additive constants in $B_{\sigma\chi}$ and $\ln \Phi_{d+1}$.
Requiring \refb{ed+1dim} to satisfy this requirement
we get the form of the solution to be:
\ben \label{esolfinform}
ds_{str,d+1}^2 &=& c_1 \left( -r^2 d\bar t^2 + {dr^2\over r^2}\right)
+ c_2 \, d\Omega_{d-3}^2
+
c_3 \left(d\sigma + c_4 \, r\,  d\bar t\right)^2 \nonumber \\
&&
+ c_5  \left(d\chi + c_6\, r\,  d\bar t\right)^2
+ c_7 \,  \left(d\sigma + c_4 \, r\,  d\bar t\right)\,
\left(d\chi + c_6\, r\,  d\bar t\right)\,
\nonumber \\
e^{2\Phi_{d+1}} &=& c_8\, \sqrt{J\over Q}\, {1\over w}\,
\nonumber \\
{1\over 2} B_{\mu\nu} dx^\mu\wedge dx^\nu
&=& c_{9} \, r\, d\sigma\wedge d\bar t + c_{10} r\, d\chi\wedge d\bar t
+ c_{11} d\sigma\wedge d\chi + \left({nw\over JQ}-1\right)^{-1/2}\,
d\sigma\wedge d\chi\, , \nonumber \\
\een
where $c_1,\ldots, c_{11}$ are some numerical constants
independent of any charges or other asymptotic data, which can
in principle be determined, up to the flat direction,
by extremizing the entropy function.
This gives the general form of the solution close to the horizon.
Since the periodicities of the $\sigma$ and $\chi$ directions, as
given in \refb{eper2}, depends on the charges, the solution
has some additional implicit dependence on the charges besides
the ones shown in \refb{esolfinform}. This can be made explicit
by rewriting the solution in terms of $(y^d,\psi,
\bar t,r)$ coordinate
system using \refb{esigmachi} such that both the compact coordinates
$y^d$ and $\psi$ have period $2\pi$.

In the spirit of the discussion at the end of section \ref{scaling} we
note that from the point of view of the near horizon geometry the
coordinate $\psi$ can be regarded as a compact direction. In that case
the entropy function $\EE$ considered here can be regarded as that of a
$(d-1)$ dimensional non-rotating black hole carrying $n$ units of
momentum and $-w$ units of winding along the $y^d$ direction and $J$
units of momentum and $Q$ units of winding along the $\psi$
direction. Thus as long as non-rotating small black holes in
$(d-1)$-dimension have finite entropy, rotating $d$-dimensional black
holes also have finite entropy.\footnote{For example if we add to the
action a $(d-1)$-dimensional Gauss-Bonnet term then the resulting
non-rotating small black holes acquire a finite entropy
\cite{Sen:2005kj}.}  Furthermore if the entropy of non-rotating small
black holes in $(d-1)$ dimension agrees with the microscopic entropy,
the constant $C$ is equal to $4\pi$.  This in turn will imply that the
entropy of the rotating small black rings in $d$ dimension also agrees
with the corresponding statistical entropy.

In this context it is also worth emphasizing that if we were studying a
small black hole with four charges in $(d-1)$ dimensions instead of a
ring in $d$-dimensions, then the near horizon geometry does contain two
arbitrary parameters whose values need to be determined from the
knowledge of the asymptotic values of the moduli field.  In particular
if we carry out an analysis analogous to that of section \ref{scaling},
we shall find that even in the scaling region the supergravity solution
continues to depend on a pair of asymptotic moduli.

\section{Discussion}
\label{sec:disc}

In this paper we studied the near-horizon geometry of the small black
ring carrying charge quantum numbers $(n,J,w,Q)$
from two different viewpoints. First we examined the full
singular black ring solution of the supergravity theory describing a
rotating spinning string and identified a scaling region where, in a
suitable coordinate system, the solution
ceased to depend on the asymptotic moduli and its dependence on the
various charges appear in a fashion such that higher derivative
corrections are insensitive to the charges.
This allowed us to express the
$\alpha'$-corrected solution in terms of a set of universal functions
independent of any parameters. The entropy computed from the
$\alpha'$-corrected solution was found to have the form $C\sqrt{nw-JQ}$ 
where $C$
is a numerical constant that cannot be computed in the absence of a
complete knowledge of all the $\alpha'$ corrections. Even if the
$\alpha'$ correction to the effective action is known, in this approach
it would be
a highly non-trivial task to actually find the solution to the equations
of motion and calculate the coefficient $C$
from the $\alpha'$-corrected action. Nevertheless the
result for the entropy found in this approach
is consistent with the result for the statistical entropy of the same
system, given by $4\pi\sqrt{nw-JQ}$.

In the second approach we focussed our attention on the near horizon
geometry instead of examining the full solution. Assuming that the
near horizon geometry has an enhanced $SO(2,1)$ symmetry besides
the manifest rotational isometries of the solution,
we can write down the general form of the solution in
terms of a set of constant parameters, and the entropy is obtained by
extremizing the entropy function with respect to these parameters.
Using the various known scaling properties
of the $\alpha'$-corrected effective action we then determined the
dependence of the entropy obtained this way on the charges, and
arrived at the same answer $C\sqrt{nw-JQ}$ for some constant $C$.
Again computation of the constant $C$ requires knowledge of the
$\alpha'$-corrected action, but in this case once we know the
action there is a simple algorithm to compute the entropy without
having to solve any differential equations.

In principle extremization of the entropy function also determines
the parameters characterizing the near horizon geometry. In practice
however this is plagued by the problem that the entropy function
relevant for this problem has two flat directions, and hence the
extremization condition does not determine the solution uniquely.
Thus which member of this two parameter family appears as the actual
near horizon geometry depends on the asymptotic data. Nevertheless
by combining the information about the asymptotic data from the
first approach with the requirement of enhanced $SO(2,1)$ symmetry
of the near horizon geometry we can determine the near horizon
geometry in terms of the charges and a few (presently
unknown) numerical constants.

Clearly the most important open problem is to find the constant $C$.
An insight into this problem can be gained from the observation that
this constant is the same as what appears in the expression for the
entropy of a non-rotating small black hole in one less dimension.
Thus agreement between the microscopic and macroscopic entropy for
non-rotating small black hole in $(d-1)$-dimension would also imply
agreement between microscopic and macroscopic entropy of a rotating
small black ring in $d$-dimensions. At present however concrete
analysis of this constant $C$ has been performed only in the case of
four dimensional small black holes \cite{Iizuka:2005uv}. Initial
studies of these black holes were based on keeping only a small
subset of higher derivative corrections to the effective action,
{\it e.g.\/}\ the F-type terms \cite{Dabholkar:2004yr} or the
Gauss-Bonnet term \cite{Sen:2005kj}, yielding the same answer
$C=4\pi$. However later a general procedure for analyzing these
black holes was developed by Kraus and Larsen \cite{Kraus:2005vz}
where, based on the assumption that the $AdS_2$ and the $S^1$ factor
of the near horizon geometry combine to form an $AdS_3$ space, they
were able to relate the coefficient $C$ to the coefficients of the
gauge and gravitational Chern-Simons terms in the action. Since
these coefficients are known exactly, $C$ also can be calculated
exactly. This yielded the same answer $C=4\pi$. In principle it
should be possible to generalize the results of Kraus and Larsen to
higher dimensional black holes, but so far this has not been done.

\section*{Acknowledgments}

N.I. would like to thank all his friends in India, especially the
string theory group at TIFR, where he shared really fantastic and fun
time together with the people there by discussing string theory, and
where part of this work was done.
M.S. would like to thank TIFR and MIT CTP where part of this work was
done.
A.D., N.I. and M.S. would also
like to thank the Aspen Center for Physics 
where part of this 
work was
done.
The work of N.I. was supported in part by the National Science
Foundation under Grant No.\ PHY99-07949.  M.S. was supported in part by
Department of Energy grant DE-FG03-92ER40701 and a Sherman Fairchild
Foundation postdoctoral fellowship.

\appendix

\section{Supergravity Small Black Rings in General Dimensions}
\label{sec:SBR_gen_dim}

In this appendix we shall construct small black ring solutions in
heterotic string theory with $d$ non-compact space-time dimensions,
describing a rotating elementary string.
The same solution in a different U-duality frame was derived in
\cite{Emparan:2001ux} as supergravity supertubes.

Consider heterotic string theory in $\bbR_t\times \bbR^{d-1}\times
S^1\times T^{9-d}$ with $4\le d\le 9$.  We denote the coordinates of
$\bbR_t$, $\bbR^{d-1}$ and $S^1$ by $t$, $\xv=(x^1,\dots,x^{d-1})$ and
$x^d$ respectively, and the coordinate radius of the $x^d$ direction by
$R_d$.  We would like to study the geometry of a small black ring
sitting in the noncompact $d$-dimensional space $\bbR_t\times
\bbR^{d-1}$. As in section \ref{scaling}, we shall regard this as a
solution in the $(d+1)$-dimensional theory obtained by dimensional
reduction of the ten dimensional heterotic string theory on
$T^{9-d}$. Thus we shall use the $(d+1)$-dimensional fields to express
the solution. This in particular will require us to take the solution to
be independent of the coordinates of $T^{9-d}$; this is done by
`smearing' the ten dimensional solution along $T^{9-d}$.

In \cite{Callan:1995hn,Dabholkar:1995nc}, a large class of
supergravity solutions were derived which correspond to a
fundamental string with an arbitrary left-moving traveling wave on
it, $\xv=\Fv(t-x^d)$, where $\Fv=(F_1,\dots,F_{d-1})$ are arbitrary
functions.
In \cite{Lunin:2001fv} (see also \cite{Mathur:2005zp}), the situation
where a fundamental string is wrapping $-w$ ($w\gg 1$) times around the
$x^d$ direction and carrying $n$ ($n\gg 1$) units of momentum along the
$x^d$ direction was considered in the particular case of $d=5$.  It was
argued there that in such a situation the supergravity description is
obtained by smearing the solution of \cite{Dabholkar:1995nc} in the
$x^d$ direction.  The smeared solution can be compactified on the $x^d$
direction, giving a five-dimensional solution with an arbitrary profile
of the fundamental string, $\xv=\Fv(v)$, in the noncompact $\bbR^4$.

This construction of \cite{Lunin:2001fv} can be straightforwardly
generalized to arbitrary $d$.  Namely, a solution of the
$(d+1)$-dimensional supergravity equations of motion, describing a
fundamental string wrapping $-w$ ($w\gg 1$) times around the $x^d$
direction, carrying $n$ ($n\gg 1$) units of momentum along the $x^d$
direction, and having an arbitrary shape in the noncompact $\bbR^{d-1}$
direction parametrized by the profile function $\xv=\Fv(v)$ ($0\le v\le
L$), is given by:
\begin{eqnarray}
 ds^2_{str,d+1}&=&
  f_f^{-1}[-(dt-A_i dx^i)^2+(dx^d-A_i dx^i)^2+(f_p-1)(d t-dx^d)^2]
  +d\xv_{d-1}^2\nonumber\label{metric_SBR_gen}\\
 e^{2\Phi_{d+1}}&=& g^2\, f_f^{-1},\qquad
 B_{td}= -(f_f^{-1}-1),\qquad
  B_{ti}=-B_{di}=f_f^{-1}\, A_i,
\end{eqnarray}
where $i=1,2,\dots,d-1$, and \footnote{The relation to the harmonic
functions in \cite{Lunin:2001fv, Mathur:2005zp} is $f_f=H^{-1}$,
$f_p=K+1$.  }
\begin{eqnarray}
 f_f(\xv)&=&1+{Q_f\over L}\int_0^L {dv\over |\xv-\Fv(v)|^{d-3}},\qquad
 f_p(\xv)=1+{Q_f\over L}\int_0^L {|\dot\Fv(v)|^2dv\over |\xv-\Fv(v)|^{d-3}},
 \nonumber\\
 A_i(\xv)&=&-{Q_f\over L}\int_0^L {\dot F_i(v)dv\over |\xv-\Fv(v)|^{d-3}}.
 \label{f_def_gen_dim}
\end{eqnarray}
The dot denotes derivative with respect to $v$, and $L=2\pi w R_d$.  We
also define
\begin{eqnarray}
 \label{edefqp}
Q_p\equiv{Q_f\over L}\int_0^L
{|\dot\Fv(v)|^2dv}\, .
\end{eqnarray}
For large $|\xv|$, $f_f-1$ and $f_p-1$ fall off as $Q_f/|{\bf
x}|^{d-3}$ and $Q_p/|\xv|^{d-3}$ respectively.  By computing the
flux of the gauge fields associated with $G_{d\mu}$ and
$B_{d\mu}$ at infinity, one finds that the relations between $Q_{f},Q_p$
and quantized charges $n,w$ are
\begin{eqnarray}
 Q_f&=&{16\pi G_d R_d\over (d-3)\Omega_{d-2}\alpha'}w,
  \qquad
 Q_p={16\pi G_d\over (d-3)\Omega_{d-2}R_d}n,
 \label{rel_QN_gen}
\end{eqnarray}
where $\Omega_D$ is the area of $S^D$ and $G_d$ is the $d$-dimensional
Newton constant:
\begin{eqnarray} \label{edefnewton}
 16\pi G_d&=&{16\pi G_{d+1}\over 2 \pi R_d}
  ={(2\pi)^{d-3} g^2\ap^{(d-1)/2}\over R_d}.
\end{eqnarray}
In order to arrive at \refb{rel_QN_gen}, \refb{edefnewton} we have used
the fact that in the absence of higher derivative corrections, which are
irrelevant in the asymptotic region, the action has the form given in
\refb{etree}--\refb{esugra}.  By examining the asymptotic form of the
metric and the results of \cite{Myers:1986un} one also sees that the
angular momentum associated with the solution in the $x^i$-$x^j$ plane
is given by
\begin{eqnarray}
\label{eangular}
J_{ij} = \frac{(d-3)\Omega_{d-2}}{16\pi G_d}
              \frac{Q_f}{L} \int_0^L (F_i \dot{F}_j - F_j \dot{F}_i) dv
              \, .
\end{eqnarray}

Before considering the small black ring solution, let us first consider
the case with a circular profile:
\begin{eqnarray}
 \Fv=\Fv^{(0)},\qquad
  \left\{
  \begin{array}{l}
   F_1^{(0)}+iF_2^{(0)}=R e^{i\omega v},\\[.5ex]
    F_3^{(0)}=\cdots=F_{d-1}^{(0)}=0,
  \end{array}
  \right.
  \qquad
 \omega={2\pi Q\over L}={Q\over w R_d}.
 \label{F=F0}
\end{eqnarray}
This corresponds to a fundamental string which winds $Q$ times along the
ring of radius $R$ in the $x^1$-$x^2$ plane
\cite{Lunin:2001fv,Lunin:2002iz}.  Introducing the coordinate system
$(s,\psi,w,\vec\xi\,)$ by
\begin{eqnarray} \label{exyz1}
 x^1&=&s\cos\psi,\quad x^2=s\sin\psi, \nonumber\\
 x^3&=&w\, \xi^1,\quad x^4=w\, \xi^2,\quad \dots,\quad x^{d-1} =w\, \xi^{d-3}\qquad
  \sum_{a=1}^{d-3} (\xi^a)^2=1,
\end{eqnarray}
the harmonic functions in \eqref{f_def_gen_dim} are computed as
\begin{eqnarray}
 f_f &=& 1 + Q_f
            (s^2+w^2+R^2)^{-{d-3\over 2}}
        {}_2 F_1(\tfrac{d-3}{4},\tfrac{d-1}{4}; 1; \tfrac{4R^2s^2}{(s^2+w^2+R^2)^2}),
        \nonumber\\
 f_p&=&1+Q_p
            (s^2+w^2+R^2)^{-{d-3\over 2}}
        {}_2 F_1(\tfrac{d-3}{4},\tfrac{d-1}{4}; 1; \tfrac{4R^2s^2}{(s^2+w^2+R^2)^2}),
        \label{fffpA}\\
 A_\psi&=&-(\tfrac{d-3}{2})q R^2 s^2
            (s^2+w^2+R^2)^{-{d-1\over  2}}
        {}_2 F_1(\tfrac{d-1}{4},\tfrac{d+1}{4}; 2; \tfrac{4R^2s^2}{(s^2+w^2+R^2)^2}),
        \nonumber
\end{eqnarray}
where we have defined
\begin{eqnarray}
 q\equiv Q_f\omega \label{edefofq}
\end{eqnarray}
and ${}_2F_1(\alpha,\beta;\gamma;z)$ denotes the
hypergeometric function.  For odd $d$, the hypergeometric functions in
\eqref{fffpA} can be written as rational functions, while for even $d$
they involve elliptic integrals.
Furthermore, from \eqref{edefqp},
\begin{eqnarray}
 \label{eQq0}
 Q_p=Q_fR^2\omega^2.
\end{eqnarray}
Using \refb{edefofq} and the last equation of \eqref{F=F0}, one finds
\begin{eqnarray}
 q={16\pi G_d\over (d-3)\Omega_{d-2}\alpha'}Q
 \, .\label{rel_qn_gen}
\end{eqnarray}
Moreover, from \refb{eangular} one finds that the solution carries an
angular momentum $J=Q R^2/\alpha'$ in the $x^1$-$x^2$ plane. This gives
\begin{eqnarray}
 \label{erdef}
 R^2=\alpha'{J\over Q}\, .
\end{eqnarray}
{}From \eqref{rel_QN_gen},
\eqref{eQq0}, \eqref{erdef},
and the last
equation of \eqref{F=F0}, we obtain
\begin{eqnarray}
 JQ=nw,\label{regge_micro0}
\end{eqnarray}
{\it i.e.\/}, the circular configuration \eqref{F=F0} saturates the
Regge bound.

Now let us proceed to construct the small black ring solution.  This can
be done by taking the profile function to be
\cite{Balasubramanian:2005qu}
\begin{eqnarray}
 \Fv=\Fv^{(0)}+\delta\Fv,\qquad
 \label{F=F0+dF}
\end{eqnarray}
where $\delta\Fv$ describes fluctuations around ${\bf F}^{(0)}$, whose
detailed form is irrelevant as long as it satisfies certain conditions
to be explained below.  As the simplest example, take $\delta \Fv$ 
to be
\begin{eqnarray}
 \delta F_1+i\,\delta F_2&=a \, e^{i(\nu v+b)},\label{dF_eg}
\end{eqnarray}
where we eventually take the limit
\begin{eqnarray}
 {a\over R}\to 0, \qquad
  {\nu\over \omega}\to \infty, \qquad
  a\nu:\mbox{ fixed.}
  \label{dFcond}
\end{eqnarray}
In other words,
$\delta \Fv$ of \eqref{dF_eg} represents a very small-amplitude
($a\ll R$), high-frequency ($\nu\gg \omega$) fluctuation.  Because of
the first condition in \eqref{dFcond}, $\Fv$ in the denominators in the
integrand of \eqref{f_def_gen_dim} can be replaced by $\Fv^{(0)}$.  Now
expand this denominator as
\begin{eqnarray}
 |\xv-\Fv^{(0)}|^{-(d-3)}
  &=& [s^2+w^2+R^2-2sR\cos(\omega v-\psi)]^{-{d-3\over 2}}
  \nonumber\\
  &=&(s^2+w^2+R^2)^{-{d-3\over 2}}
  \sum_{k=0}^\infty{1\over k!}{\Gamma({d-3\over 2}+k)\over\Gamma({d-3\over 2})}
  \left[{2sR\cos(\omega v-\psi)\over s^2+w^2+R^2}\right]^{k}.\label{jhyc11Nov06}
\end{eqnarray}
On the other hand, in the numerator, {\it e.g.\/}\ for $f_p$, we have
\begin{eqnarray}
 |\dot\Fv|^2
  &=&
 |\dot\Fv^{(0)}+\delta\dot\Fv|^2
 =
  R^2\omega^2+a^2\nu^2+2R\omega a\nu \cos[(\omega-\nu)v-b].\label{ismz11Nov06}
\end{eqnarray}
When we multiply \eqref{jhyc11Nov06} and \eqref{ismz11Nov06}, and
integrate it over $v$, there will be nonvanishing contributions only
when the frequencies of the cosines in \eqref{jhyc11Nov06} and
\eqref{ismz11Nov06} cancel each other, which happens only for $k\gtrsim
{\nu\over \omega}$.  Since ${\nu\over \omega}\to\infty$ in the limit
\eqref{dFcond}, in fact there is no contribution from the last term in
\eqref{ismz11Nov06}. Similarly, there is no contribution from $\delta
\Fv$ to $A_i$ in \eqref{f_def_gen_dim}; only $\Fv^{(0)}$ contributes.

At the end of the day, the only effect of introducing the fluctuation
\eqref{dF_eg} is to change eq.\ \eqref{eQq0} to
\begin{eqnarray}
 \label{eQq}
 Q_p=Q_f(R^2\omega^2+a^2\nu^2),
\end{eqnarray}
whereas other expressions \eqref{fffpA}, 
\eqref{rel_qn_gen} and
\eqref{erdef} are unchanged.  These give the supergravity small black
ring solution we were after.  Note that eqs.\
\eqref{rel_QN_gen}, \eqref{eQq},
\eqref{erdef}, and the last equation of \eqref{F=F0}
now imply the Regge bound,
\begin{eqnarray}
  JQ<nw.\label{regge_micro}
\end{eqnarray}
Even if one considers more complicated fluctuations than \eqref{dF_eg}
by taking linear combinations of many modes in all the $x^i$
directions ($1\le i\le (d-1)$),
the above results remain unchanged as long as the condition
\eqref{dFcond} is met for each mode, except that the $a^2\nu^2$ term in
\eqref{eQq} will be replaced by a sum over the contribution
from all the modes.
The fact that the resulting solution \eqref{fffpA} is insensitive to the
precise form of the fluctuation $\delta \Fv$ is the reflection of the
fact that this supergravity small black ring represents all the
underlying microstates whose entropy is given by \eqref{earbdim}.

\bigskip

Although in \eqref{fffpA} we presented the small black ring solution in
the $(s,\psi,w,\vec\xi\,)$ coordinate system, it is more convenient for
the analysis in the main text to go to the coordinate system
$(y,\psi,x,\vec\xi\,)$ defined by
\begin{eqnarray} \label{exyz2}
 s&=&{\sqrt{y^2-1}\over x-y}R,\quad w={\sqrt{1-x^2}\over x-y}R,\qquad
  -1\le x\le 1,\quad -\infty<y\le -1.
\end{eqnarray}
In terms of these coordinates, the harmonic functions \eqref{fffpA}
become
\begin{eqnarray}
 f_f &=& 1 + \tfrac{Q_f}{R^{d-3}} (\tfrac{x-y}{-2y})^{(d-3)/ 2}
        {}_2 F_1(\tfrac{d-3}{4},\tfrac{d-1}{4}; 1; 1-\tfrac{1}{y^2}),
        \nonumber\\
 f_p &=& 1 + \tfrac{Q_p}{R^{d-3}} (\tfrac{x-y}{-2y})^{(d-3)/ 2}
        {}_2 F_1(\tfrac{d-3}{4},\tfrac{d-1}{4}; 1; 1-\tfrac{1}{y^2}),
        \label{fffpA_xy}\\
 A_\psi&=&-(\tfrac{d-3}{2})\tfrac{q}{R^{d-5}}
            \tfrac{(y^2-1)(x-y)^{(d-5)/2}}{(-2y)^{(d-1)/2}}
        {}_2 F_1(\tfrac{d-1}{4},\tfrac{d+1}{4}; 2; 1-\tfrac{1}{y^2}),
        \nonumber
\end{eqnarray}
and the $(d-1)$-dimensional flat metric $d\xv_{d-1}^2$
can be written as
\begin{eqnarray}
 d\xv_{d-1}^2
 &=& {R^2\over (x-y)^2}
  \left[
   {dy^2\over y^2-1}+(y^2-1)d\psi^2+{dx^2\over 1-x^2}
   +(1-x^2)d\Omega_{d-4}^2
  \right]. \label{enewcoord}
\end{eqnarray}

Eqs.\ \refb{metric_SBR_gen}, \refb{fffpA_xy} and \refb{enewcoord},
together with the definitions of various parameters given in
\refb{rel_QN_gen}, \refb{edefnewton}, \refb{rel_qn_gen} and
\refb{erdef}, describe the supergravity small black ring solution.  Note
that once the solution has been obtained this way, we can forget about
how it was constructed, and simply analyze its properties by treating
this as a singular solution of the supergravity equations of
motion. This is the view point we have adopted in section \ref{scaling}.


\begin{thebibliography}{99}

\bibitem{Maldacena:1997de}
  J.~M.~Maldacena, A.~Strominger and E.~Witten,
  ``Black hole entropy in M-theory,''
  JHEP {\bf 9712}, 002 (1997)
  [arXiv:hep-th/9711053].

  \bibitem{Behrndt:1998eq}
  K.~Behrndt, G.~Lopes Cardoso, B.~de Wit, D.~Lust, T.~Mohaupt and W.~A.~Sabra,
  ``Higher-order black-hole solutions in N = 2 supergravity
  and Calabi-Yau
  string backgrounds,''
  Phys.\ Lett.\ B {\bf 429}, 289 (1998)
  [arXiv:hep-th/9801081].

  \bibitem{LopesCardoso:1998wt}
  G.~Lopes Cardoso, B.~de Wit and T.~Mohaupt,
  ``Corrections to macroscopic supersymmetric black-hole entropy,''
  Phys.\ Lett.\ B {\bf 451}, 309 (1999)
  [arXiv:hep-th/9812082].

  \bibitem{LopesCardoso:1999cv}
  G.~Lopes Cardoso, B.~de Wit and T.~Mohaupt,
  ``Deviations from the area law for supersymmetric black holes,''
  Fortsch.\ Phys.\  {\bf 48}, 49 (2000)
  [arXiv:hep-th/9904005].

\bibitem{Mohaupt:2000mj}
  T.~Mohaupt,
  ``Black hole entropy, special geometry and strings,''
  Fortsch.\ Phys.\  {\bf 49}, 3 (2001)
  [arXiv:hep-th/0007195].

\bibitem{LopesCardoso:1999ur}
  G.~Lopes Cardoso, B.~de Wit and T.~Mohaupt,
  ``Macroscopic entropy formulae and non-holomorphic corrections for
  supersymmetric black holes,''
  Nucl.\ Phys.\ B {\bf 567}, 87 (2000)
  [arXiv:hep-th/9906094].

\bibitem{LopesCardoso:2000qm}
  G.~Lopes Cardoso, B.~de Wit, J.~Kappeli and T.~Mohaupt,
  ``Stationary BPS solutions in N = 2 supergravity with R**2 interactions,''
  JHEP {\bf 0012}, 019 (2000)
  [arXiv:hep-th/0009234].

\bibitem{Cardoso:2000fp}
  G.~L.~Cardoso, B.~de Wit, J.~Kappeli and T.~Mohaupt,
  ``Examples of stationary BPS solutions in N = 2 supergravity theories  with
  R**2-interactions,''
  Fortsch.\ Phys.\  {\bf 49}, 557 (2001)
  [arXiv:hep-th/0012232].

\bibitem{Dabholkar:2004yr}
  A.~Dabholkar,
  ``Exact counting of black hole microstates,''
  Phys.\ Rev.\ Lett.\  {\bf 94}, 241301 (2005)
  [arXiv:hep-th/0409148].

\bibitem{Dabholkar:2004dq}
  A.~Dabholkar, R.~Kallosh and A.~Maloney,
  ``A stringy cloak for a classical singularity,''
  JHEP {\bf 0412}, 059 (2004)
  [arXiv:hep-th/0410076].

\bibitem{Sen:2004dp}
A.~Sen, ``How does a fundamental string stretch its
  horizon?',''
  JHEP {\bf 0505}, 059 (2005) [arXiv:hep-th/0411255].

  \bibitem{Hubeny:2004ji}
  V.~Hubeny, A.~Maloney and M.~Rangamani,
  ``String-corrected black holes,''
  JHEP {\bf 0505}, 035 (2005)
  [arXiv:hep-th/0411272].

\bibitem{Behrndt:2005he}
  K.~Behrndt, G.~Lopes Cardoso and S.~Mahapatra,
 ``Exploring the relation between 4D and 5D BPS solutions,''
  Nucl.\ Phys.\ B {\bf 732}, 200 (2006)
  [arXiv:hep-th/0506251].


\bibitem{Dabholkar:1989jt}
  A.~Dabholkar and J.~A.~Harvey,
  ``Nonrenormalization of the Superstring Tension,''
  Phys.\ Rev.\ Lett.\  {\bf 63}, 478 (1989).

\bibitem{Sen:1995in}
  A.~Sen,
  ``Extremal black holes and elementary string states,''
  Mod.\ Phys.\ Lett.\ A {\bf 10}, 2081 (1995)
  [arXiv:hep-th/9504147].

\bibitem{Peet:1995pe}
  A.~W.~Peet,
   ``Entropy And Supersymmetry Of D-Dimensional
   Extremal Electric Black Holes
  Versus String States,''
  Nucl.\ Phys.\ B {\bf 456}, 732 (1995)
  [arXiv:hep-th/9506200].

\bibitem{Sen:1997is}
  A.~Sen,
  ``Black holes and elementary string states in
  N = 2 supersymmetric string
  theories,''
  JHEP {\bf 9802}, 011 (1998)
  [arXiv:hep-th/9712150].

\bibitem{Sen:2005kj}
  A.~Sen,
  ``Stretching the horizon of a higher dimensional small black hole,''
  arXiv:hep-th/0505122.

\bibitem{Kraus:2005vz}
  P.~Kraus and F.~Larsen,
  ``Microscopic black hole entropy in theories with higher derivatives,''
  arXiv:hep-th/0506176.

\bibitem{Kraus:2005zm}
  P.~Kraus and F.~Larsen,
  ``Holographic gravitational anomalies,''
  JHEP {\bf 0601}, 022 (2006)
  [arXiv:hep-th/0508218].

\bibitem{Kraus:2006wn}
  P.~Kraus,
  ``Lectures on black holes and the AdS(3)/CFT(2) correspondence,''
  arXiv:hep-th/0609074.

\bibitem{Sen:2005pu}
  A.~Sen,
  ``Black holes, elementary strings and holomorphic anomaly,''
  JHEP {\bf 0507}, 063 (2005)
  [arXiv:hep-th/0502126].

\bibitem{Dabholkar:2005by}
  A.~Dabholkar, F.~Denef, G.~W.~Moore and B.~Pioline,
  ``Exact and asymptotic degeneracies of small black holes,''
  JHEP {\bf 0508}, 021 (2005)
  [arXiv:hep-th/0502157],

\bibitem{Dabholkar:2005dt}
  A.~Dabholkar, F.~Denef, G.~W.~Moore and B.~Pioline,
  ``Precision counting of small black holes,''
  JHEP {\bf 0510}, 096 (2005)
  [arXiv:hep-th/0507014].



\bibitem{Bena:2004tk}
  I.~Bena and P.~Kraus,
  ``Microscopic description of black rings in AdS/CFT,''
  JHEP {\bf 0412}, 070 (2004)
  [arXiv:hep-th/0408186].

\bibitem{Bena:2005ay}
  I.~Bena and P.~Kraus,
  ``Microstates of the D1-D5-KK system,''
  Phys.\ Rev.\ D {\bf 72}, 025007 (2005)
  [arXiv:hep-th/0503053].

\bibitem{Gaiotto:2005gf}
  D.~Gaiotto, A.~Strominger and X.~Yin,
  ``New connections between 4D and 5D black holes,''
  JHEP {\bf 0602}, 024 (2006)
  [arXiv:hep-th/0503217].

\bibitem{Elvang:2005sa}
  H.~Elvang, R.~Emparan, D.~Mateos and H.~S.~Reall,
  JHEP {\bf 0508}, 042 (2005)
  [arXiv:hep-th/0504125].

\bibitem{Gaiotto:2005xt}
  D.~Gaiotto, A.~Strominger and X.~Yin,
  ``5D black rings and 4D black holes,''
  arXiv:hep-th/0504126.

\bibitem{Bena:2005ni}
  I.~Bena, P.~Kraus and N.~P.~Warner,
  Phys.\ Rev.\ D {\bf 72}, 084019 (2005)
  [arXiv:hep-th/0504142].

\bibitem{Iizuka:2005uv}
  N.~Iizuka and M.~Shigemori,
  ``A note on D1-D5-J system and 5D small black ring,''
  arXiv:hep-th/0506215.

\bibitem{Elvang:2004rt}
  H.~Elvang, R.~Emparan, D.~Mateos and H.~S.~Reall,
  ``A supersymmetric black ring,''
  Phys.\ Rev.\ Lett.\  {\bf 93}, 211302 (2004)
  [arXiv:hep-th/0407065].

\bibitem{Bena:2004de}
  I.~Bena and N.~P.~Warner,
  ``One ring to rule them all ... and in the darkness bind them?,''
  arXiv:hep-th/0408106.

\bibitem{Elvang:2004ds}
  H.~Elvang, R.~Emparan, D.~Mateos and H.~S.~Reall,
  ``Supersymmetric black rings and three-charge supertubes,''
  Phys.\ Rev.\ D {\bf 71}, 024033 (2005)
  [arXiv:hep-th/0408120].

\bibitem{Gauntlett:2004qy}
  J.~P.~Gauntlett and J.~B.~Gutowski,
  ``General concentric black rings,''
  Phys.\ Rev.\ D {\bf 71} (2005) 045002
  [arXiv:hep-th/0408122].

\bibitem{Mathur:2005zp}
  S.~D.~Mathur,
  ``The fuzzball proposal for black holes: An elementary review,''
  Fortsch.\ Phys.\  {\bf 53}, 793 (2005)
  [arXiv:hep-th/0502050],

\bibitem{Mathur:2005ai}
  S.~D.~Mathur,
  ``The quantum structure of black holes,''
  Class.\ Quant.\ Grav.\  {\bf 23}, R115 (2006)
  [arXiv:hep-th/0510180].

\bibitem{Lunin:2002qf}
  O.~Lunin and S.~D.~Mathur,
   ``Statistical interpretation of Bekenstein entropy for systems with a
  stretched horizon,''
  Phys.\ Rev.\ Lett.\  {\bf 88}, 211303 (2002)
  [arXiv:hep-th/0202072].

\bibitem{Russo:1994ev}
  J.~G.~Russo and L.~Susskind,
  ``Asymptotic level density in heterotic string theory and rotating black
  holes,''
  Nucl.\ Phys.\ B {\bf 437}, 611 (1995)
  [arXiv:hep-th/9405117].

\bibitem{Palmer:2004gu}
  B.~C.~Palmer and D.~Marolf,
  ``Counting supertubes,''
  JHEP {\bf 0406}, 028 (2004)
  [arXiv:hep-th/0403025].

\bibitem{Marolf:2004fy}
  D.~Marolf and B.~C.~Palmer,
  ``Gyrating strings: A new instability of black strings?,''
  Phys.\ Rev.\ D {\bf 70}, 084045 (2004)
  [arXiv:hep-th/0404139].

\bibitem{Bak:2004kz}
  D.~Bak, Y.~Hyakutake, S.~Kim and N.~Ohta,
  ``A geometric look on the microstates of supertubes,''
  Nucl.\ Phys.\ B {\bf 712}, 115 (2005)
  [arXiv:hep-th/0407253].

\bibitem{Emparan:2001ux}
  R.~Emparan, D.~Mateos and P.~K.~Townsend,
  ``Supergravity supertubes,''
  JHEP {\bf 0107}, 011 (2001)
  [arXiv:hep-th/0106012].

\bibitem{Dabholkar:1995nc}
  A.~Dabholkar, J.~P.~Gauntlett, J.~A.~Harvey and D.~Waldram,
  ``Strings as Solitons \& Black Holes as Strings,''
  Nucl.\ Phys.\ B {\bf 474}, 85 (1996)
  [arXiv:hep-th/9511053].

\bibitem{Sen:2005wa}
  A.~Sen,
   ``Black hole entropy function and the attractor mechanism in higher
  derivative gravity,''
  JHEP {\bf 0509}, 038 (2005)
  [arXiv:hep-th/0506177].

\bibitem{Dabholkar:2005qs}
  A.~Dabholkar, N.~Iizuka, A.~Iqubal and M.~Shigemori,
  ``Precision microstate counting of small black rings,''
  Phys.\ Rev.\ Lett.\  {\bf 96}, 071601 (2006)
  [arXiv:hep-th/0511120].

\bibitem{Callan:1995hn}
  C.~G.~Callan, J.~M.~Maldacena and A.~W.~Peet,
  ``Extremal Black Holes As Fundamental Strings,''
  Nucl.\ Phys.\ B {\bf 475}, 645 (1996)
  [arXiv:hep-th/9510134].

\bibitem{Lunin:2001fv}
  O.~Lunin and S.~D.~Mathur,
  ``Metric of the multiply wound rotating string,''
  Nucl.\ Phys.\ B {\bf 610}, 49 (2001)
  [arXiv:hep-th/0105136].

\bibitem{Balasubramanian:2005qu}
  V.~Balasubramanian, P.~Kraus and M.~Shigemori,
   ``Massless black holes and black rings as effective
   geometries of the D1-D5
  system,''
  Class.\ Quant.\ Grav.\  {\bf 22}, 4803 (2005)
  [arXiv:hep-th/0508110].

\bibitem{Wald:1993nt}
R.~M.~Wald,
``Black hole entropy in the Noether charge,''
Phys.\ Rev.\ D {\bf 48}, 3427 (1993)
[arXiv:gr-qc/9307038].

\bibitem{Jacobson:1993vj}
 T.~Jacobson, G.~Kang and R.~C.~Myers,
 ``On black hole entropy,''
Phys.\ Rev.\ D {\bf 49}, 6587 (1994) [arXiv:gr-qc/9312023].

\bibitem{Iyer:1994ys}
V.~Iyer and R.~M.~Wald,
``Some properties of Noether charge and a proposal for dynamical black
hole
entropy,''
Phys.\ Rev.\ D {\bf 50}, 846 (1994)
[arXiv:gr-qc/9403028].

\bibitem{Jacobson:1994qe}
T.~Jacobson, G.~Kang and R.~C.~Myers,
``Black hole entropy in higher curvature gravity,''
arXiv:gr-qc/9502009.

\bibitem{Sen:1994eb}
  A.~Sen,
  ``Black hole solutions in heterotic string theory on a torus,''
  Nucl.\ Phys.\ B {\bf 440}, 421 (1995)
  [arXiv:hep-th/9411187].

\bibitem{unpublished}
A.~Dabholkar, N.~Iizuka, A.~Iqubal, A.~Sen
    and M.~Shigemori, unpublished.

\bibitem{Astefanesei:2006dd}
  D.~Astefanesei, K.~Goldstein, R.~P.~Jena, A.~Sen and S.~P.~Trivedi,
  ``Rotating attractors,''
  arXiv:hep-th/0606244.

  \bibitem{Sahoo:2006pm}
  B.~Sahoo and A.~Sen,
   ``alpha' corrections to extremal dyonic black holes in heterotic string
  theory,''
  arXiv:hep-th/0608182.

\bibitem{Myers:1986un}
  R.~C.~Myers and M.~J.~Perry,
  ``Black Holes In Higher Dimensional Space-Times,''
  Annals Phys.\  {\bf 172}, 304 (1986).

\bibitem{Lunin:2002iz}
  O.~Lunin, J.~M.~Maldacena and L.~Maoz,
  ``Gravity solutions for the D1-D5 system with angular momentum,''
  arXiv:hep-th/0212210.
\end{thebibliography}
\end{document}